\documentclass[journal,12pt,onecolumn,draftclsnofoot,]{IEEEtran}
\usepackage{cite}
\ifCLASSINFOpdf
    \usepackage[pdftex]{graphicx}
    \graphicspath{{./figures/}}
\else
    \usepackage[dvips]{graphicx}
    \graphicspath{{./figures/}}
\fi
\usepackage{amsmath}
\usepackage{amssymb}
\usepackage{amsthm}
\theoremstyle{definition}
\newtheorem{definition}{Definition}

\newcommand{\R}{\mathbb{R}}

\usepackage{algorithm}
\usepackage{algpseudocode}
\usepackage{array}
\usepackage{makecell}
\usepackage{multicol}
\usepackage{multirow}
\usepackage{threeparttable}
\usepackage{subfigure}
\usepackage{float}
\usepackage{stfloats}
\usepackage{url}
\usepackage{hyperref}

\newcommand{\aref}[1]{\hyperref[#1]{Appendix~\ref*{#1}}}

\usepackage[nameinlink]{cleveref}
\crefformat{equation}{(#2#1#3)}

\usepackage{xcolor}

\usepackage{soul}

\usepackage{newtxtext}

\begin{document}
\title{Rate-Adaptive Coding Mechanism for Semantic Communications With Multi-Modal Data}
% example
% \author{Michael~Shell,~\IEEEmembership{Member,~IEEE,}
%         John~Doe,~\IEEEmembership{Fellow,~OSA,}
%         and~Jane~Doe,~\IEEEmembership{Life~Fellow,~IEEE}% <-this % stops a space
% \thanks{M. Shell was with the Department
% of Electrical and Computer Engineering, Georgia Institute of Technology, Atlanta,
% GA, 30332 USA e-mail: (see http://www.michaelshell.org/contact.html).}% <-this % stops a space
% \thanks{J. Doe and J. Doe are with Anonymous University.}% <-this % stops a space
% \thanks{Manuscript received April 19, 2005; revised August 26, 2015.}}

% \author{Yangshuo~He,~\IEEEmembership{Student~Member,~IEEE}% <-this % stops a space
\author{Yangshuo~He,~Guanding~Yu,~and~Yunlong~Cai%
\thanks{The authors are with the College of Information Science \& Electronic Engineering, Zhejiang University, 38 Zheda Road, Hangzhou, China, 310027, email: \{sugarhe@zju.edu.cn, yuguanding@zju.edu.cn, ylcai@zju.edu.cn\}}
}


% \markboth{Journal of \LaTeX\ Class Files,~Vol.~14, No.~8, August~2015}%
% {Shell \MakeLowercase{\textit{et al.}}: Bare Demo of IEEEtran.cls for IEEE Journals}
\markboth{}%
{}



\maketitle

\begin{abstract}

Recently, the ever-increasing demand for bandwidth in multi-modal communication systems requires a paradigm shift. Powered by deep learning, semantic communications are applied to multi-modal scenarios to boost communication efficiency and save communication resources. However, the existing end-to-end neural network (NN) based framework without the channel encoder/decoder is incompatible with modern digital communication systems. Moreover, most end-to-end designs are task-specific and require re-design and re-training for new tasks, which limits their applications. In this paper, we propose a distributed multi-modal semantic communication framework incorporating the conventional channel encoder/decoder. We adopt NN-based semantic encoder and decoder to extract correlated semantic information contained in different modalities, including speech, text, and image. Based on the proposed framework, we further establish a general rate-adaptive coding mechanism for various types of multi-modal semantic tasks. In particular, we utilize unequal error protection based on semantic importance, which is derived by evaluating the distortion bound of each modality. We further formulate and solve an optimization problem that aims at minimizing inference delay while maintaining inference accuracy for semantic tasks. Numerical results show that the proposed mechanism fares better than both conventional communication and existing semantic communication systems in terms of task performance, inference delay, and deployment complexity.
\end{abstract}
  
% Note that keywords are not normally used for peer review papers.

% \fi
\IEEEpeerreviewmaketitle

\section{Introduction}
\label{sec:introduction}
%\IEEEpubidadjcol

Modern communication systems are developed based on the Shannon information theory \cite{information-theory} to recover transmitted messages and use bit rate or bit error rate (BER) as a key performance metric. 
With the coming era of connected intelligence \cite{6G-horizon}, transmitting the escalated amount of data becomes a huge burden on communication systems. 
Under the conventional Shannon's framework, the trending solutions are demanding wider and wider bandwidth together with higher and higher transmission power. 
However, the ever-increasing requests for communication resources will eventually face a serious bottleneck. 
Therefore, as a new paradigm shift, semantic communications \cite{semantic-communication} are investigated again to reduce the amount of transmitted data irrelevant to the final task. 

The promising benefits of semantics have inspired various researchers ever since Shannon information theory. The first measure of semantic information based on the logical probability of propositions was established by \cite{outline-semantic-theory}. \cite{towards-semantic-theory} further refined this idea by proposing a generic semantic communication model in propositional logic. However, these historical works only affect messages with logical probability, which is infeasible in reality.
Current studies on semantic communications generally focus on end-to-end learning based on deep neural networks (NNs), which leverage the power of deep learning to extract semantic information from source data and learn to minimize the negative effects caused by imperfect wireless channels. 
The core ideas of these works are training an end-to-end NN with manually added channel noise and utilizing task-specific metrics rather than bit rate or BER.
Specifically, the NN encoder extracts semantic information from data and the NN decoder utilizes the semantic information for downstream tasks, including data reconstruction and inference.
The work in \cite{JSCC-text} first adopted deep learning for joint source and channel coding (JSCC), in which text data is compressed and recovered via a stacked long short term memory (LSTM) network. 
Then, the authors in \cite{JSCC-image} followed this idea and implemented JSCC with a convolutional neural network (CNN) for image transmission.  
The transformer encoder is utilized in \cite{DeepSC} to better extract the semantic information for JSCC. 
To cope with semantic representation, the authors in \cite{IB,IB-multiuser} utilized the information bottleneck theory to minimize the size of semantic features while maximizing the relevance between data and task output.
Moreover, the authors in \cite{VQ-VAE} leveraged a vector quantized-variational autoencoder to learn a codebook of semantic features, which is shared by both transmitter and receiver. The authors in \cite{personal-saliency} further expanded upon the task-oriented concept and proposed a personalized multi-user semantic communication framework for image transmission.

On the other hand, as the emerging connected intelligence in daily life, various networked sensors collect numerous multi-modal data from the physical world, thereby driving a large amount of transmitted data \cite{multi-modal-survey}. 
In conventional communication systems, accurately transmitting numerous data would cause a bottleneck in communication resources. 
However, the aforementioned semantic-aware and task-oriented communications show their great potential for processing massive multi-modal data. 
One main task of semantic communications is to merge multi-modal data for further improving the inference performance \cite{what-is-semantic}, such as human sentiment detection, Level 5 autonomous driving, automation control in smart house, and real-time health care analysis. In such scenarios, semantic information is contained in the multi-modal data. By integrating the modality-specific features from multi-modal data and discarding redundant invariant information, the task goal of reducing communication overhead can be achieved. 
In \cite{VQA}, a multi-modal semantic communication framework named DeepSC-VQA was proposed. Based on JSCC system, DeepSC-VQA processes image and text separately, and fuses semantic information for visual question answering. 
However, existing multi-modal semantic communications face two main challenges:
\begin{enumerate}
    \item First, most existing studies on semantic communications implement conventional source and channel encoders using NNs. The output signals of such NNs are in general continuous values that are incompatible with modern digital communication systems. To address such a problem, some studies have introduced quantization techniques into NNs. The authors in \cite{DeepSC-lite} quantized the full resolution constellation, which is the output tensor of the NN, to $64$-QAM during the inference phase. Furthermore, the work in \cite{DeepJSCC-Q} adopted the constellation quantization during the feedforward process. Regarding the back-propagation, the zero derivative problem of the quantization function is solved by replacing the hard quantization symbol with a differentiable soft-max weighted symbol. However, the quantization value is no longer the optimal result of the NN's objective function. Moreover, the derivatives substitution in back-propagation is unexplainable, which limits the performance of the multi-modal fusion. 
    \item Secondly, most existing studies adopt the end-to-end NN for extracting semantic information and mitigating the effect of channel noise. 
    % The NN is only designed and optimized to extract semantic features for a specific task. When the semantic communication system is required to transmit data for another task, it is needed to modify the NN architecture and re-train the network. 
    The NN in current semantic communication studies is optimized jointly with a wireless channel for a specific task, instead of directly utilizing the existing pre-trained state-of-the-art model for this task. When the semantic communication system is required to transmit data for another task or other wireless environment, it is needed to re-design and train an NN or re-train the network with a new wireless channel using the pre-trained model.
    However, it is usually infeasible to re-train an end-to-end model in practical multi-modal scenarios from time to time. For example, in a distributed sensor network, it is unrealistic to re-train an encoder with all kinds of decoders. To handle the above limitation, the concept of multi-task learning has been introduced into semantic communications in \cite{multitask}. The authors proposed a multi-task semantic communication framework named U-DeepSC, which can serve several tasks with fixed parameters. Nevertheless, U-DeepSC can only serve tasks in a small set and thus re-training is still unavoidable for other transmission tasks. Furthermore, due to the limited resources of sensor devices, it is impractical to train complex models with a large number of parameters in multi-modal semantic communications.
\end{enumerate}

In this paper, to deal with the first challenge, we propose a distributed multi-modal semantic communication framework that incorporates the key modules in the conventional physical layer.
We utilize the NN-based semantic source encoder while retaining the conventional channel encoder. 
The semantic source encoder aims at extracting semantic information from different modality data and discarding irrelevant and redundant data. The following channel encoder protects semantic features from imperfect wireless channels. 
Unlike existing JSCC frameworks that utilize black box NNs for joint encoders, our design not only effectively corrects errors due to channel noise in a more effective and explainable manner but also can be easily implemented on well-established digital communication systems.

Based on the distributed multi-modal semantic communication framework, we further propose a rate-adaptive coding mechanism to address the second challenge. Specifically, we take semantic importance into consideration in channel coding and assign unequal error protection for different semantic information accordingly. The novelty of the proposed mechanism is twofold. First, the proposed mechanism is independent of the NN-based semantic encoder and decoder. The semantic encoder and decoder can be directly implemented with a separated pre-trained multi-modal fusion model while the aforementioned JSCC approach requires end-to-end re-training. Secondly, we leverage the robustness verification problem (RVP) \cite{robustness-verification-problem} to evaluate the effect of channel coding on semantic task performance. Given the universality of robustness bounds computation, the proposed mechanism can be applied to all types of NNs and model inputs. Therefore, our proposed design is general to various multi-modal fusion tasks that are realized by separable pre-trained NN models.

To summarize, our main contributions are as follows:
\begin{enumerate}
    \item \textbf{Distributed Multi-Modal Semantic Communication Framework:} We propose a practical distributed multi-modal semantic communication framework that incorporates channel encoders to be compatible with modern communication systems. The NN-based semantic source encoder can be implemented with existing pre-trained models for different tasks without re-training since the encoder is designed independently of the physical layer. Additionally, the proposed framework supports unequal error protection for different modalities according to their semantic significance, which exploits the semantic diversity among different modalities.
    \item \textbf{Theoretical Analysis of Semantic Importance:} We utilize RVP to characterize the relation between the input perturbation for different modalities and the distortion of semantic output. We consider the input distortion as channel noise and further evaluate the semantic inference performance in realistic wireless channels. Based on the noise sensitivity of different modalities, we define the semantic importance of different modalities for semantic tasks.
    \item \textbf{Algorithm Development for Rate-Adaptive Coding Mechanism:} We present a rate-adaptive coding mechanism that jointly considers semantic information and channel conditions to unequally assign coding rates for different modalities according to their semantic significance. Specifically, we formulate and solve an optimization problem in multi-modal semantic communications, where the coding rates of different modalities are optimized to minimize the inference delay under semantic task performance constraints.
\end{enumerate}

The rest of this paper is organized as follows: The distributed multi-modal semantic communication framework is introduced in \autoref{sec:system_model}.
\autoref{sec:semantic_importance_analysis} analyzes the semantic importance of different modalities for unequal protection.
\autoref{sec:rate_adaptive_coding_mechansim} further develops the algorithm for the rate-adaptive coding mechanism.
Numerical results are presented in \autoref{sec:simulation_results} to illustrate the effectiveness of the proposed rate-adaptive coding mechanism. Finally, conclusions are drawn in \autoref{sec:conclusion}.

\section{Multi-Modal Semantic Communication Framework}
\label{sec:system_model}

\begin{figure}[htbp]
    \centering
    \includegraphics[width=0.9\textwidth]{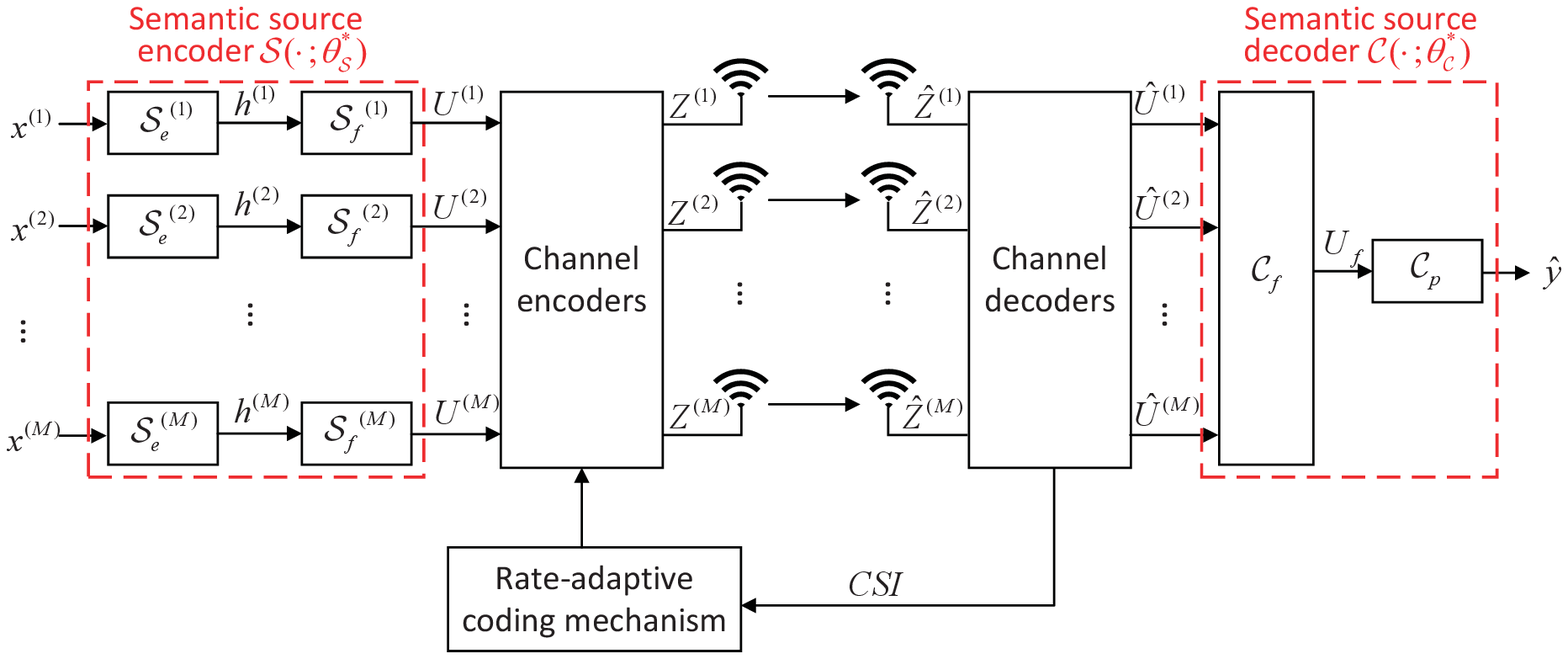}
    \caption{The distributed multi-modal fusion semantic communication framework.}
    \label{fig:distributed_multimodal_fusion}
    \vspace{-0.5cm}
\end{figure}

In this section, we first introduce the distributed multi-modal fusion semantic communication framework for wireless communication systems. As shown in \autoref{fig:distributed_multimodal_fusion}, we mainly focus on multi-modal semantic communications in a distributed co-inference system. 
% A well-trained multi-modal fusion model $\mathcal{F}:\mathcal{X}\to\mathcal{Y}$ is utilized to perform inference tasks, e.g., sentiment analysis and emotion recognition.
To carry out semantic tasks such as sentiment analysis and emotion recognition, we employ a well-trained multi-modal fusion model, denoted as $\mathcal{F}:\mathcal{X}\to\mathcal{Y}$.
This model $\mathcal{F}:\mathcal{X}\to\mathcal{Y}$ is split into semantic source encoders $\mathcal{S}=\left\{\mathcal{S}^{(1)},\mathcal{S}^{(2)},\cdots,\mathcal{S}^{(M)}\right\}$ and a semantic source decoder $\mathcal{C}$, which are separately deployed at edge devices and the server, respectively.
Specifically, the model takes $M$ different modalities collected from edge devices as input $X=\left\{x^{(1)}, x^{(2)}, \cdots,x^{(M)}\right\}$. The semantic source encoder for the $m$-th modality extracts the semantic information as
\begin{equation}
    U^{(m)} = \mathcal{S}^{(m)}\left(x^{(m)};\theta_{\mathcal{S}^{(m)}}\right),
\end{equation}
where $U^{(m)}\in \R^{d^{(m)}}$ is the semantic information of the $m$-th modality and $\mathcal{S}^{(m)}\left(\cdot;\theta_{\mathcal{S}^{(m)}}\right)$ is the semantic encoder of the $m$-th modality with learnable parameters $\theta_{\mathcal{S}^{(m)}}$. 
The process of semantic encoding can be decomposed into two consecutive procedures: modality embedding and feature extraction. 
Modal embedding is a feature extraction technique that represents modality data as low-dimensional vectors.
In practice, the observed multi-modal data are typically large inputs with different formats, such as text, images, and audio. Therefore, the modality-specific module $\mathcal{S}_e^{(m)}$ is employed to convert raw data $x^{(m)}$ into the corresponding modal embedding $h^{(m)}$. For instance, BERT is a recently developed textual embedding model, while CNN is widely used for visual embeddings.
However, the embeddings of various modalities would still consume a significant communication overhead. Hence, a feature extractor $\mathcal{S}_f^{(m)}$ is used to further process and compress the modal embedding to output semantic information of the $m$-th modality.

After that, the rate-adaptive coding mechanism utilizes semantic information to evaluate the semantic importance of different modalities. Combining with the channel state information (CSI) feedback, the transmission rate for each modality is optimized to minimize the inference delay while maintaining semantic task performance. 
In the rate-adaptive coding mechanism, the semantic features are quantized into a binary bit stream for physical layer transmission. To simplify the following analysis, we normalize semantic features to $[0, 1]$ and quantize the floating-point precision features into $B$-bit fixed-point binary representation, denoted by $\Tilde{U}^{(m)}$. 
The bit-level semantic features are then encoded and modulated into transmission symbols $Z^{(m)}$, which are transmitted to the server through wireless channels.
At the receiver, the channel decoder decodes the received symbols $\hat{Z}^{(m)}$ into $\hat{U}^{(m)}$ and feeds back CSI to the transmitter for the rate-adaptive coding mechanism.
Due to imperfect wireless transmission, $\hat{U}^{(m)}$ is generally not identical with $\Tilde{U}^{(m)}$. 

At the server, a feature-level fusion mechanism $\mathcal{C}_f$ fuses $\hat{\mathbf{U}}^M=\left\{\hat{U}^{(1)},\hat{U}^{(2)}\cdots,\hat{U}^{(M)}\right\}$ into a joint representation $U_f$ for downstream semantic tasks. Since the semantic information contained in $M$ modalities is closely related, $\mathcal{C}_f$ is capable of learning the cross-modality representation based on the significance of each modality. Finally, a semantic task function $\mathcal{C}_p$ takes the fusion output to conduct a specific task. Note that, the semantic decoder consists of both multi-modal fusion and task-specific functions. The distorted output of the semantic source decoder is given by
\begin{equation}
    \label{equ:semantic_decoder}
    \hat{y} = \mathcal{C}\left(\hat{\mathbf{U}}^M;\theta_{\mathcal{C}}\right),
\end{equation}
where $\theta_{\mathcal{C}}$ is the learnable parameters of semantic decoder $\mathcal{C}$.

The NN-based semantic encoder shows great potential to extract semantic information from the complex semantic source. However, the end-to-end NN design lacks error protection ability against channel noise and is incompatible with modern bit-level communication systems. 
The proposed distributed multi-modal semantic communication framework only utilizes NN-based semantic encoders for source coding and maintains the conventional channel encoder for error protection. Based on this, we can further propose a rate-adaptive coding mechanism that jointly considers the semantic encoder and conventional channel coding to minimize inference delay while maintaining semantic task accuracy. 
In the following, we present the theoretic analysis and algorithm development for the proposed rate-adaptive coding mechanism in detail.

\section{Semantic Importance Analysis}
\label{sec:semantic_importance_analysis}

In the multi-modal fusion problem, different modalities contribute differently to the semantic task. The observed fact reveals that the distortion tolerance for each modality varies. Specifically, significant modalities are more sensitive to noise, but the trivial ones allow relatively larger distortion without affecting the overall inference performance of the semantic task. 
As a result, the unequal error protection scheme based on semantic importance can be applied in multi-modal fusion semantic communications. 
In this section, we first present a brief introduction to RVP. Then we theoretically investigate the robustness bound of the semantic output and formulate the semantic importance.

\subsection{Introduction of RVP}
Intuitively, if the $m$-th modality is semantically important to the semantic task, the perturbation of this modality feature will cause a considerable semantic output distortion and even severely degrade the inference performance. To investigate the relation between input distortion and output distortion, we leverage the RVP in \cite{robustness-verification-problem}. Although the NNs are trained to behave correctly for unseen inputs, it is observed that the performance of NNs can be severely degraded by slightly perturbed inputs. Therefore, RVP has been proposed to provide formal guarantees for the behavior of NNs. An NN is robust if it produces similar outputs for every perturbed input within a small bound. 
The authors in \cite{robustness-verification} introduced the concept of robustness bounds in multi-layer perceptron (MLP), which represent the lower and upper distortion bounds of the neuron output with bounded perturbed inputs. 
RVP has also been utilized to analyze the robustness of various NNs, such as CNN \cite{robustness-CNN}, RNN \cite{robustness-RNN}, LSTM \cite{robustness-LSTM}, and Transformer \cite{robustness-Transformer}. Therefore, the following robustness analysis can be generally applied to different neural network tasks with various architectures, and is not just limited to the multi-modal semantic task in our simulations.

In communication systems, the imperfect wireless channel would cause distortions in received signals and degrade the performance of neural networks. Therefore, the RVP is adopted in \cite{robustness-GNN} to verify the robustness of graph neural networks in a decentralized inference scenario. Similarly, we model the channel noise as the bounded perturbation and leverage RVP in the rate-adaptive coding mechanism.
Specifically, the received semantic features $\mathbf{U}^M$ are distorted by the random channel noise and thus the semantic output $\hat{y}$ also deviates from the original true output $y$.
To ensure the robustness of the semantic task, we utilize RVP to compute the distortion bound of the semantic output, which is a function of input semantic features and distortion bounds of each modality. The perturbation bounds of different modalities are related to transmission error probabilities that are determined by specific channel coding rates. Consequently, based on this we are able to analyze the effect of channel coding rate on semantic task performance. 

\subsection{Robustness Bound in Multi-Modal Semantic Communications}
\label{sec:RVP}
We assume that the perturbed semantic feature $\hat{U}^{(m)}$ is within an $\mathcal{L}_p$ ball, i.e. $\hat{U}^{(m)}\in\mathbb{B}_p\left(U^{(m)}, \Delta^{(m)}\right)$, where $\mathbb{B}_p(x, \Delta)=\left\{\hat{x}:\|\hat{x}-x\|_p\leq \Delta\right\}$ and $\|\cdot\|_p$ denotes the $\mathcal{L}_p$ norm. Given the perturbed input $\hat{\mathbf{U}}^M$, the distorted semantic output is bounded by $\mathcal{C}^L$ and $\mathcal{C}^U$, i.e., $\mathcal{C}^L\preceq \hat{y} \preceq \mathcal{C}^U$. We consider the range of distortion as the robustness bound on the output value. Therefore, the maximum robustness bound over all perturbed input is given by 
\begin{equation}
    \label{equ:robustness_bound}
    \gamma\left(\mathcal{C}, \mathbf{U}^M, \mathbf{\Delta}^M\right) = \max_{\hat{\mathbf{U}}^M} \mathcal{C}^U- \min_{\hat{\mathbf{U}}^M} \mathcal{C}^L,
\end{equation}
where $\mathbf{\Delta}^M=\left\{\Delta^{(1)},\Delta^{(2)},\cdots,\Delta^{(M)}\right\}$ is the maximum distortion bound of each semantic feature. To compute the lower and upper bounds of the semantic output, we adopt the linear-relaxation framework \cite{robustness-verification}, which models the bound as a linear function and utilizes linear relaxation for non-linear operations in the NN. By propagating the bound parameters over the NN, we can obtain the function of linear lower and upper bounds of the output w.r.t the semantic features.

There are several methods available for computing the bound parameters in an NN. 
Without loss of generality, we represent the NN as a computational graph, similar to TensorFlow and PyTorch. This representation enables general robustness bound analysis for various NN architectures.
Moreover, for the NN without recurrent structure \footnote{Analysis of the robustness bounds of non-recurrent neural networks such as RNN and LSTM can also refer to previous studies \cite{robustness-RNN} and \cite{robustness-LSTM}.}, the corresponding computational graph can be considered as a directed acyclic graph (DAG) $G_\mathcal{C}=(\mathbf{V},\mathbf{E})$, where $\mathbf{V}=\{v_1,v_2,\cdots,v_n\}$ is the set of computation nodes in $\mathcal{C}$ and $\mathbf{E}=\left\{\left(v_i,v_j\right)\right\}$ is the set of node pairs representing that $v_i$ is the input node of $v_j$.
Each computation node $v_i$ represents an NN operation $\mathbf{h}_i = h_i\left(u(i)\right)\in \R^{d_i}$, where $u(i)$ is the set of parent nodes for node $v_i$. For simplicity, we represent the computation function recursively as a function of the input feature $\mathbf{h}_i = h_i\left(\hat{\mathbf{U}}^M\right)$. According to the attribute of DAG, we can recursively compute the linear lower and upper bounds of the output node. Before that, we first give the definition of linear bounds on computation nodes.
\begin{definition}[Linear bounds on computation nodes \cite{robustness-verification}]
\label{def:bound}
For computation node $v_i$ in DAG, we define two linear functions $h_i^L,h_i^U:\mathbb{R}^{d_1}\to \mathbb{R}^{d_i}$, $\mathbf{h}_i^L=h_i^L\left(\hat{\mathbf{U}}^M\right)=\mathbf{A}_i^L\hat{\mathbf{U}}^M + \mathbf{b}_i^L,\mathbf{h}_i^U=h_i^U\left(\hat{\mathbf{U}}^M\right)=\mathbf{A}_i^U\hat{\mathbf{U}}^M + \mathbf{b}_i^U$, such that $\mathbf{h}_i^L\leq \mathbf{h}_i \leq \mathbf{h}_i^U$. $\mathbf{A}_i^{L}\mathbf{A}_i^{U}\in\mathbb{R}^{d_i\times d_1},\mathbf{b}_i^{L},\mathbf{b}_i^{U}\in\mathbb{R}^{d_i}$ are parameters of linear bounds given by the feed-forward function $F_i$ of node $v_i$ which takes parameters for every parent nodes $v_j\in u(i)$ as input, i.e., $\left(\mathbf{A}_i^{L},\mathbf{A}_i^{U},\mathbf{b}_i^{L},\mathbf{b}_i^{U}\right) = F_i\left(\left\{\mathbf{A}_j^{L},\mathbf{A}_j^{U},\mathbf{b}_j^{L},\mathbf{b}_j^{U})\right\}_{j\in u(i)}\right)$.
\end{definition}

To gain more insight into the feed-forward function $F_i$, some detailed examples of $F_i$ in various neural network architectures can be found in previous studies \cite{robustness-verification,robustness-CNN,robustness-RNN,robustness-LSTM,robustness-Transformer}.
According to \autoref{def:bound}, the bounds of semantic output are also the bounds of the output node $v_n$. Therefore, the lower and upper bounds of the semantic output for perturbed input $\hat{\mathbf{U}}^M$ are respectively given by 
\begin{equation}
    \label{equ:output_bound}
    \left\{
        \begin{array}{c}
            \mathcal{C}^L=\mathbf{h}_n^L=\mathbf{A}_n^L\hat{\mathbf{U}}^M + \mathbf{b}_n^L, \\
            \mathcal{C}^U=\mathbf{h}_n^U=\mathbf{A}_n^U\hat{\mathbf{U}}^M + \mathbf{b}_n^U,
        \end{array}
    \right.
\end{equation}
where the parameters $\mathbf{A}_n^{L},\mathbf{A}_n^{U}$, $\mathbf{b}_n^{L}$, and $\mathbf{b}_n^{U}$ can be recursively computed from input nodes by the virtue of DAG. Note that the parameters for input node $v_1$ are $\mathbf{A}_1^L=\mathbf{A}_1^U=\mathbf{I},\ \mathbf{b}_1^L=\mathbf{b}_1^U=\mathbf{0}$. Next, we compute the maximum upper bound and minimum lower bound for the semantic output over every perturbed input in order to obtain the maximum robustness bounds $\gamma$, which is the maximum distortion range of $\hat{y}$. The maximum upper bound over all perturbed input $\hat{\mathbf{U}}^M$ can be computed as
\begin{align}
    \label{equ:max_upper_bound}
    \begin{split}
        \max_{\hat{\mathbf{U}}^M} \mathcal{C}^U & = \max_{\hat{\mathbf{U}}^M} \mathbf{A}_n^U\hat{\mathbf{U}}^M + \mathbf{b}_n^U \\
        & = \sum_{m=1}^M \max_{\hat{U}^{(m)}\in\mathbb{B}_p(U^{(m)}, \Delta^{(m)})} \mathbf{A}_n^{(m),U}\hat{U}^{(m)} + \mathbf{b}_n^{(m),U} \\
        & = \sum_{m=1}^M \Delta^{(m)} \max_{J\in\mathbb{B}_p(\mathbf{0},1)} \mathbf{A}_n^{(m),U}J + \mathbf{A}_n^{(m),U}U^{(m)} + \mathbf{b}_n^{(m),U} \\
        & = \sum_{m=1}^M \Delta^{(m)} \left\|\mathbf{A}_n^{(m),U}\right\|_q + \mathbf{A}_n^{(m),U}U^{(m)} + \mathbf{b}_n^{(m),U},
    \end{split}
\end{align}
where $\|\cdot\|_q$ is the dual norm of $\mathcal{L}_p$ norm meaning that $q$ satisfies $1/p + 1/q = 1$.
The detailed proof can be found in \aref{apx:dual_norm}. 

The minimum lower bound can be computed in a similar way. In our distributed multi-modal semantic communication framework, the distortion of semantic features is caused by the imperfect wireless channel. To jointly design channel coding and semantic encoder/decoder, we only need to consider the first term which is related to channel noise. Accordingly, the maximum robustness bound of the semantic output in \cref{equ:robustness_bound} can be further formulated as
\begin{equation}
    \label{equ:output_bound_result}
    \gamma\left(\mathcal{C}, \mathbf{U}^M, \mathbf{\Delta}^M\right) = \sum_{m=1}^M \Delta^{(m)}\left(\left\|\mathbf{A}_n^{(m),U}\right\|_q+\left\|\mathbf{A}_n^{(m),L}\right\|_q\right).
\end{equation}

Since the perturbation is due to wireless channels with random noise, the input distortion is positively related to the transmission error probability, i.e. $\Delta^{(m)}\propto\epsilon$. Moreover, the error probability can be expressed as $\epsilon=Q(\varphi)$, where $Q(\cdot)$ is the Gaussian Q-function and $\varphi$ is the signal-to-noise-ratio (SNR). If the SNR is large enough, the error probability is approximated to zero, i.e. $Q(\varphi)\to 0,\ \varphi\gg 0$. Therefore, in the high SNR scenario, the maximum robustness bound of semantic outputs could be approximated to zero which indicates that the distortion bound is tight.

\subsection{Definition of Semantic Importance}

It can be seen from \cref{equ:output_bound_result} that the maximum robustness bound of semantic outputs is a weighted summation of input distortion bounds, where the weight is the norm of the lower and upper bound parameters. Clearly, given the same input perturbation, the modalities with larger parameters would significantly enlarge the robustness bound of the semantic output. Therefore, these modalities take a crucial role in the semantic task and thus we can formulate the semantic importance of the $m$-th modality as
\begin{equation}
    \label{equ:semantic-importance}
    \kappa^{(m)} = \left\|\mathbf{A}_n^{(m),U}\right\|_q+\left\|\mathbf{A}_n^{(m),L}\right\|_q.
\end{equation}

In order to enhance the robustness of semantic inference, we should pay more attention to those semantically important modalities. Intuitively, we can minimize the perturbation of more important modalities to reduce the distortion bound in \cref{equ:output_bound_result}. More specifically, we can use lower coding rates for the transmission of important modalities to mitigate the effect of imperfect wireless channels. Meanwhile, distortions on those less semantically significant modalities would have relatively less impact on the semantic output by the virtue of small weights. As a result, high coding rates could be assigned for those unimportant modalities to improve communication efficiency. In the next section, we will discuss the unequal error protection for the rate-adaptive coding mechanism in detail.

\section{Rate-Adaptive Coding Mechanism}
\label{sec:rate_adaptive_coding_mechansim}

It is observed that different modalities may play different roles in the semantic task. To improve communication efficiency, we introduce unequal error protection to the joint source and channel coding design. Specifically, the proposed rate-adaptive coding mechanism first characterizes the semantic importance of all $M$ modalities and then encodes each modality feature with different error protection abilities. For simplicity, we only consider the channel coding rate $R^{(m)}$, which is the most important error protection factor for channel code. In other words, under the premise of successful semantic task execution, the proposed mechanism allocates each modality a coding rate for improving communication efficiency. In this section, we first formulate the rate-adaptive coding problem and then develop an efficient algorithm to solve it.

\subsection{Problem Formulation}
As aforementioned, the goal of the proposed rate-adaptive coding mechanism is to maximize communication efficiency under given semantic task robustness requirements. Considering that each edge device is capable of sensing and processing only one type of modality, the inference latency can be expressed as
\begin{equation}
    T=\max_{m}\{T_m^C + T_m^U\},
\end{equation}
where $T_m^C$ is the calculation latency for the $m$-th edge device and $T_m^U$ is the upload latency for the $m$-th edge device to transmit modality features. We assume that the calculation delay is neglectable \footnote{We neglect the calculation delay in our channel transmission problem because it is not the focus of our research and can be handled by adding a constant to the optimal solution.}. We simplify the transmission delay  for the $m$-th modality as $T_m^U = D^{(m)}/R^{(m)}$, where $D^{(m)}$ is the data size of semantic features and $R^{(m)}$ is the transmission rate. Therefore, the inference latency of the semantic task can be expressed as $T=\max_{m}\{D^{(m)}/R^{(m)}\}$.

In practice, modalities have different data sizes, as well as coding rates, which are determined by semantic importance. As a result, the transmission latency for different modalities varies greatly. In this case, the inference of semantic tasks may be severely delayed if the transmission time of one modality is too high. Meanwhile, other devices must wait for the straggler modality, which causes a considerable waste of time and resources. To achieve high inference efficiency, we formulate the rate-adaptive coding mechanism as an optimization problem whose objective is to minimize the overall inference delay, as
\begin{align}
    \label{equ:problem_P1}
    \begin{split}
        \mathcal{P}1: \min_{\{R^{(m)}\}_M} & \max_{m} \frac{D^{(m)}}{R^{(m)}}, \\
        \text{s.t.} \quad & \gamma\left(\mathcal{C}, \mathbf{U}^M, \mathbf{\Delta}^M\right) \leq \Delta_0, \\
        & R^{(m)} \leq C^{(m)}-\sqrt{\frac{V^{(m)}}{L}} Q^{-1}(\epsilon^{(m)}) \log_2 e+\frac{\log L}{L}, \quad \forall m=1,2, \cdots, M, \\
        & \Delta^{(m)} \leq \sum_{j=1}^{\epsilon^{(m)}B} 2^{-j}, \quad \forall m=1,2, \cdots, M, 
    \end{split}
\end{align}
where $\epsilon^{(m)}$ is the error probability, $Q^{
-1}(\cdot)$ is the inverse of $Q(x)=\int_x^\infty \text{exp}\left(-t^2/2\right)/\sqrt{2\pi}dt$, $L$ is the blocklength of the channel code, $\Delta_0$ is the bound of semantic task distortion and $C^{(m)},V^{(m)}$ are respectively the channel capacity and dispersion, given by
\begin{equation}
    \left\{
    \begin{array}{c}
        C^{(m)} = \frac{1}{2}\log_2\left(1 + \varphi^{(m)}\right), \\
        V^{(m)} = 1 - \left(1 + \varphi^{(m)}\right)^{-2}.
    \end{array}
    \right.
\end{equation}

The received SNR is given by $\varphi^{(m)}=|h^{(m)}|^2P^{(m)}/P_{\text{noise}}^{(m)}$, where $h^{(m)}$ is the channel gain, $P^{(m)}$ is the transmit power, and $P_{\text{noise}}^{(m)}=\sigma^{(m)}W$ is the noise power. The robustness of the semantic task is formulated as the first constraint. According to \cite{finite-length-channel-coding}, the second constraint specifies the maximal achievable rate with error probability $\epsilon$. The last constraint connects the numerical distortion of each modality and the transmission error probability. It comes from the fact that bit errors in the most significant bits would cause the maximum distortion for $B$-bit fixed-point quantization. 

\subsection{Problem Transformation}

The problem in \cref{equ:problem_P1} is not easy to solve, since the $Q$ function does not have an operational closed-form. In the following, we approximate it into a convex one. First, we utilize the approximation of $Q$ function in \cite{page-approximation}, as
\begin{equation}
    \label{equ:approx}
    Q(z) \approx \frac{1}{1 + \text{exp}(2az)},
\end{equation}
where $a = \sqrt{2/\pi}$. In the proposed framework, NNs are robust to distortions caused by the imperfect wireless channel. Thus, it is possible to allow for a relatively higher error probability aiming at a high transmission rate, which corresponds to the scenario that the approximation gap of \cref{equ:approx} is negligible. Then, we further look into the last constraint in \cref{equ:problem_P1}. Under the fixed-point quantization, the maximum distortion given $K$ error bits is $\sum_{k=1}^K 2^{-k}$, which indicates the errors on the $K$ most significant bits. We denote the indices of error bits as $e_K = \{i_1,i_2,\cdots,i_K\}$. The distortion caused by an error on the $i_k$-th bit is denoted as $d(i_k) = 2^{-i_k}$. We further assume that the probability of bit flip satisfies i.i.d. uniform distribution. Thus the probability of $K$ bits error is $\Pr\{e_K\} = \tbinom{B}{K}^{-1}$, where $\tbinom{B}{K} = B!/(K!(B-K)!)$. Based on this, the expectation of numerical distortion with $K$ error bits is calculated as
\begin{equation}
    \text{E}\{d_K\} = \sum_{e_K}\Pr\{e_K\}\sum_{k=1}^Kd(i_k) = \frac{\tbinom{B-1}{K-1}}{\tbinom{B}{K}}\sum_{l=1}^B2^{-l} = \frac{K}{B}\left(1-2^{-B}\right).
\end{equation}
Therefore, we can approximate the distortion constraint in \cref{equ:problem_P1} to $\Delta^{(m)}\leq \epsilon^{(m)}\left(1-2^{-B}\right)$. The approximation gap tends to be zero and negligible when high precision quantization is used.

For convenience, we can equivalently rewrite the objective function as a formulation of maximizing the minimum rate $\max_{\{R^{(m)}\}_M} \min_m R^{(m)}/D^{(m)}$.
To solve the minimizing maximum problem in \cref{equ:problem_P1}, we further transform the problem equivalently by introducing a new variable $\tau = \min_m R^{(m)}/D^{(m)}$ and adding a constraint of $\tau \leq R^{(m)}/D^{(m)}$. Considering the optimal case that the rate and distortion for each modality would reach their maximum, we could take equal signs for the last two constraints. By combining all the aforementioned steps, $\mathcal{P}1$ can be approximately converted to
\begin{align}
    \label{equ:problem_P2}
    \begin{split}
        \mathcal{P}2:  \min_{\{R^{(m)}\}_M,\tau} & -\tau, \\
        \text{s.t.} \quad & \tau \leq \frac{R^{(m)}}{D^{(m)}}, \quad \forall m=1,2, \cdots, M, \\
        & \sum_{m=1}^M\frac{\left(1-2^{-B}\right)\kappa^{(m)}}{1 + \text{exp}\left[\sqrt{\frac{8L}{\pi V^{(m)} \log_2^2 e}}(C^{(m)}+\frac{\log L}{L}-R^{(m)})\right]} \leq \Delta_0. 
    \end{split}
\end{align}

We can easily prove that $\mathcal{P}2$ is a standard convex optimization problem. The detailed proof can be found in \aref{apx:convex}. In the case of the optimal solution $\tau^*$, it is apparent that the distortion of each modality reaches the maximum bound and all modalities share the same transmission delay. Therefore, the constraints in $\mathcal{P}2$ can take the equal sign when $\tau=\tau^*$. We use the Lagrange multiplier method to find that the optimal solution $\tau^*$ satisfies
\begin{equation}
    \label{equ:optimal_solution}
    f(\tau^*) =\sum_{m=1}^{M} \frac{a^{(m)}}{1+\text{exp}\left[k^{(m)}(b^{(m)}-D^{(m)} \tau^*)\right]}-\Delta_{0}=0,
\end{equation}
where $a^{(m)} \triangleq \left(1-2^{-B}\right)\kappa^{(m)}$, $b^{(m)} \triangleq C^{(m)}+\log L/L$, and $k^{(m)} \triangleq \sqrt{8L/\pi V^{(m)} \log_2^2 e}$. It is difficult to find a closed-form solution of summation of exponents in \cref{equ:optimal_solution}. Consequently, we can solve $\mathcal{P}2$ using the bisection method as shown in \autoref{alg:bisection}. 
% TODO删减对分搜索
\begin{algorithm}[htb]
    \caption{Numerical Algorithm for $\mathcal{P}2$}
    \label{alg:bisection}
    \begin{algorithmic}[1]
        \Require
            tolerance, $TOL$; maximum iterations, $N_0$ 
        \Ensure 
            optimal code rate, $\{R^{(m)*}\}_M$
        \State Set counter $i\gets 1$
        \State $a\gets 0 ,\quad b\gets \frac{\max _{m} C^{(m)}}{\min _{m} D^{(m)}},\quad FA\gets-\Delta_0,\quad f(\tau)\gets \sum_{m=1}^{M} \frac{a^{(m)}}{1+\mathrm{exp}\left[k^{(m)}\left(b^{(m)}-D^{(m)} \tau\right)\right]}-\Delta_{0}$
        \State $\tau^*\gets \mathrm{bisection}(a,b,FA,f)$
        % \While{$i\leq N_0$}
        %     \State $\tau\gets a + (b-a)/2,\quad F\gets \sum_{m=1}^{M} \frac{a^{(m)}}{1+\mathrm{exp}\left[k^{(m)}\left(b^{(m)}-D^{(m)} \tau\right)\right]}-\Delta_{0}$
        %     \If{$F=0$ or $(b-a)/2<TOL$}
        %         \State $\tau^*\gets \tau$
        %         \State \Return $\tau^*$
        %     \EndIf
        %     \State $i\gets i + 1$
        %     \If{$FA\cdot F>0$}
        %         \State $a\gets \tau,\quad FA\gets F$
        %     \Else
        %         \State $b\gets \tau$
        %     \EndIf
        % \EndWhile
        \State $R^{(m)*}\gets D^{(m)}\tau^*$
    \end{algorithmic}
\end{algorithm}

By observation, we can easily obtain the range of the optimal solution to the optimization problem, i.e., $\tau^*\in\left(0, \min_m \left(C^{(m)}+\log L / L\right)/D^{(m)}\right]$.
When $\tau = 0$, the transmission rate $R^{(m)}=0$ which indicates the infinite length repetition coding. In this situation, $f(0)=-\Delta_0$ since the error probability tends to be zero. 
In the case of the upper bound, the error probability $\epsilon^{(m)}=0.5$, which is equivalent to random transmission. We assume $\tau_{\text{max}} = \left(C^{(1)}+\log L/L\right)/D^{(1)}$ without loss of generality. Then, we have $f(\tau_{\text{max}}) = A^{(1)}/2 + \sum_{m=2}^{M}a^{(m)}/\left(1+e^{r^{(m)}}\right) -\Delta_0$ from \cref{equ:optimal_solution}, where $r^{(m)} = k^{(m)}\left(b^{(m)}-b^{(1)}D^{(m)}/D^{(1)}\right)\geq 0$. To satisfy the condition in the bisection method, we should emphasize the constraint on choosing the maximum robustness bound of semantic output, i.e., $\Delta_0\leq 0.5\sum_{m=1}^Ma^{(m)}$. Combining it with \cref{equ:output_bound_result}, this selected constraint indicates that the maximum distortion bounds of all modality features are $\Delta^{(m)}\approx 0.5$. 

We further look into the equation in \cref{equ:optimal_solution} and find that the variable $\tau$ can be regarded as the minimum weighted coding rate over all modalities. The inference delay of the semantic task decreases with the increase of the weighted coding rate. Meanwhile, the increasing weighted coding rate enlarges the modality distortion bound, which eventually deteriorates the semantic task performance. Therefore, the optimal solution $\tau^*$ balances the tradeoff between the inference delay and inference accuracy.

\subsection{Algorithm Development}

Thus far, we have characterized the semantic importance of different modalities by leveraging RVP to evaluate the input distortion bounds. Furthermore, we have analyzed the semantic task performance under unequal error protection, based on which we formulate and solve an optimization problem to balance the tradeoff between the inference delay and inference accuracy. The optimal transmission rate for each modality is derived from the aforementioned problem. In the following, we summarize the proposed algorithm for the rate-adaptive coding mechanism, as shown in \autoref{alg:rate-adaptive}. 

\begin{figure}[htbp]
    \centering
    \includegraphics[width=0.5\textwidth]{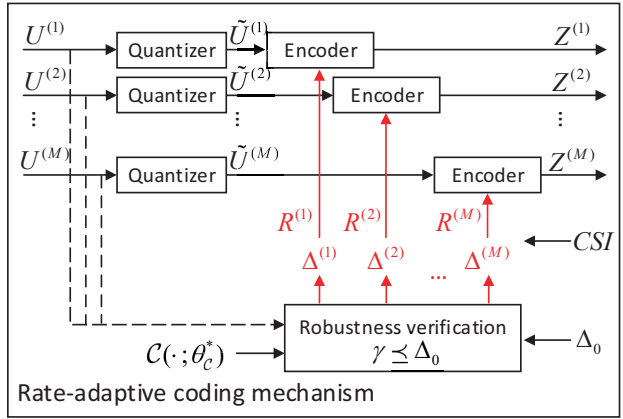}
    \caption{The proposed rate-adaptive coding mechanism.}
    \label{fig:rate_adaptive_coding_mechanism}
    \vspace{-0.5cm}
\end{figure}

\begin{algorithm}[htb]

    \caption{Rate-Adaptive Coding Mechanism}
    \label{alg:rate-adaptive}
    \begin{algorithmic}[1]
        \Require
            modality features, $\mathbf{U}^M$; semantic source decoder, $\mathcal{C}\left(\cdot;\theta_{\mathcal{C}}^*\right)$, output distortion bound $\Delta_0$, channel state information $CSI$
        \Ensure
            transmission symbols, $\mathbf{Z}^M$
        \State Recursively compute parameters of linear lower and upper bounds, $\mathbf{A}_n^{L/U},\mathbf{b}_n^{L/U}\gets \text{RVP}\left(\mathbf{U}^M, \mathcal{C}\right)$
        \State Compute the maximum robustness bound of semantic output, $\gamma\gets \sum_{m=1}^M \Delta^{(m)}\kappa^{(m)}$
        \State Estimate SNR $\varphi\gets CSI$
        \State Solve the optimal transmission rate in $\mathcal{P}2$, $\{R^{(m)*}\}_M\gets \mathcal{P}2(\gamma, \Delta_0, \varphi)$
        \For{modality $m$}
            \State Quantize modality feature into $B$-bit representation, $\Tilde{U}^{(m)}\gets \text{Quantize}\left(U^{(m)}\right)$
            \State Determine code rate and modulation from transmission rate, $R_c^{(m)*}, Q_m^{(m)*}\gets R^{(m)*}$
            \State Channel coding and modulation, $Z^{(m)}\gets \text{Encoder}\left(\Tilde{U}^{(m)}, R_c^{(m)}, Q_m^{(m)*}\right)$
        \EndFor
        \State $\mathbf{Z}^M\gets \left\{Z^{(1)}, Z^{(1)}, \cdots, Z^{(M)}\right\}$
    \end{algorithmic}
\end{algorithm}
As shown in \autoref{fig:rate_adaptive_coding_mechanism}, the rate-adaptive coding mechanism adaptively assigns different coding rates for different semantic features based on both communication efficiency and semantic task performance. We assume that the rate-adaptive coding mechanism has the knowledge of the well-trained semantic source decoder $\mathcal{C}\left(\cdot;\theta_{\mathcal{C}}^*\right)$. In each transmission process, the semantic source encoders map different modality data into modality features $\mathbf{U}^M$. With a restriction on the output distortion bound $\Delta_0$ and the semantic features $\mathbf{U}^M$, the RVP finds the optimal transmission rates for different modalities. Specifically, based on the modality features and NN architecture, we recursively compute the parameters of linear lower and upper bounds as in \cite{robustness-verification} and thus we compute the maximum robustness bound over all perturbed input $\gamma$. After that, we evaluate the SNR according to the CSI, which is estimated by the server. Using the numerical method in \autoref{alg:bisection}, we solve the optimization problem $\mathcal{P}2$ to find the optimal transmission rates for all modalities. For each modality, we determine the appropriate coding rate and modulation order corresponding to the optimal transmission rate. Then, the full precision semantic feature $U^{(m)}$ is quantized into a $B$-bit fixed-point binary sequence $\tilde{U}^{(m)}$, which is finally encoded and modulated into transmission symbols $Z^{(m)}$.

\subsection{Discussion}
In the following, we analyze the algorithmic complexity and compare several key features to illustrate the different between our proposed mechanism and the JSCC approach.

It can be seen that the rate-adaptive coding mechanism could be implemented using the existing pre-trained model without re-training. As a result, we only need to consider the complexity of the optimization problem which consists of RVP and the bisection method. The complexity of RVP is the complexity of semantic source decoder $\mathcal{O}(N_\mathcal{C})$, since the RVP computation process is equivalent to the feedforward process of the NN. Moreover, the complexity of the bisection method can be computed as $\mathcal{O}\left(-\log_2TOL\right)$, which is determined by the tolerance. Therefore, the complexity of the rate-adaptive coding mechanism can be written as
\begin{equation}
    \mathcal{O}\left(N_\mathcal{C} - \log_2TOL\right).
\end{equation}

Next, we analyze the complexity of the JSCC approach. Since it is impossible to transmit the full precision signals generated by the JSCC semantic encoder in practical systems, we assume that the binarized signals are encoded using channel codes and modulated to transmission symbols. Therefore, we neglect the complexity of channel coding in the following comparison. We further assume that there exist well-trained NN models for different multi-modal semantic tasks. In this case, re-training the model for end-to-end inference is necessary to deploy the JSCC approach. Thus, the complexity of JSCC can be computed as $\mathcal{O}\left(N_{\text{epoch}}\cdot N_{\text{sample}}\cdot N\right)$, where $N_{\text{epoch}}$ is the number of training epochs to converge, $N_{\text{sample}}$ is the number of samples in the training set, and $N$ is the complexity of the NN feedforward process.

Apparently, both $N_{\text{epoch}}$ and $N_{\text{sample}}$ are considerably large which means that it would take a massive effort in re-training or fine-tuning the NNs. On the other hand, the proposed rate-adaptive coding mechanism requires no re-training and thus has a much lower complexity. Hence, it is generally easier to deploy in a realistic system. The detailed complexity comparison of a specific task will be presented in \autoref{sec:performance_analysis}.

To further compare the proposed rate-adaptive coding mechanism in terms of several key features in \autoref{tab:feature_comparison}.

\begin{table}[htb]
    % \linespread{1.0}
    \caption{Comparison of the JSCC Approach and the Proposed Mechanism.}
    \begin{center}
        \begin{tabular}{|m{0.5\textwidth}<{\centering}|m{0.2\textwidth}<{\centering}|m{0.2\textwidth}<{\centering}|}
            \hline
                                        & JSCC & Ours  \\
            \hline
            Use of neural networks & 
            Yes  & Yes   \\
            \hline
            Extraction of semantic information & 
            Yes  & Yes   \\
            % \hline
            % Neural network deployment & 
            % Edge devices and the server  & Edge devices and the server   \\
            \hline
            Directly use a pre-trained state-of-the-art model & 
            No  & Yes   \\
            \hline
            Need for re-training a model with a wireless channel layer & 
            Yes  & No   \\
            \hline
            Training process optimization & 
            Yes  & No   \\
            \hline
            Inference process optimization & 
            No  & Yes   \\
            \hline
            Consideration of channel condition & 
            Implicit  & CSI feedback in optimization \\
            \hline
            Consideration of semantic significance of different modalities & 
            Implicit  & Explainable RVP method  \\
            \hline
            Method to mitigate the impact of imperfect wireless channel & 
            Deep learning method  & Channel coding with unequal error protection  \\
            \hline
            Task performance & 
            % Over fitting & Close to error-free transmission performance  \\
            Medium & Good \\
            \hline
            Implementation complexity & 
            High  & Low  \\
            \hline
        \end{tabular}
        \label{tab:feature_comparison}
    \end{center}
    \vspace{-1cm}
\end{table}

\section{Simulation Results}
\label{sec:simulation_results}
In this section, we illustrate the performance of the distributed multi-modal fusion model and validate the effectiveness of the proposed rate-adaptive coding mechanism. We demonstrate the effectiveness of the proposed scheme by comparing it with the conventional transmission mechanism and the NN-based JSCC semantic communication system.

\subsection{Simulation Setup}
\subsubsection{Datasets}
In the simulation, we implement the semantic task with multi-modal sentiment analysis (MSA) \cite{MSA}. We conduct simulation on two benchmark datasets that provide word-aligned utterances with textual, visual, and acoustic signals.
\begin{itemize}
    \item CMU-MOSI \cite{MOSI} is the first opinion-level annotated dataset for sentiment and subjectivity analysis in videos. The dataset is a collection of 2,199 online video clips expressing some opinions, each of which is annotated with a sentiment score in the range of $[-3, 3]$. The dataset is rigorously annotated with labels for subjectivity, sentiment intensity, per-frame and per-opinion annotated visual features, and per-milliseconds annotated audio features.
    \item CMU-MOSEI \cite{MOSEI} is the largest MSA dataset to date and is an improvement over MOSI. The dataset consists of more than 23,400 annotated utterances from more than 1,000 gender-balanced YouTube speakers. 
\end{itemize}
Predicting the sentiment score for each utterance can be considered as a regression task. Therefore, we adopt the mean absolute error (MAE) and Pearson correlation (Corr) as performance metrics.
\subsubsection{Neural Network Model}
We implement the distributed multi-modal fusion network based on a state-of-the-art (SOTA) model for MSA, named MISA \cite{MISA}, and we could directly use the pre-trained model provided in \cite{MISA}. To satisfy the separability of the NN in our distributed multi-modal semantic communication framework, some complicated fusion designs in the network are discarded and the detailed implementation is shown in \autoref{tab:distributed_multimodal_fusion}. We utilize the BERT-base-uncased pre-trained model \cite{BERT} for textual embedding $\mathcal{S}_e^{(1)}$, which is followed by a fully connected (FC) layer textual encoder $\mathcal{S}_f^{(1)}$. The model contains 12 Transformer encoder layers. For visual embedding $\mathcal{S}_e^{(2)}$, we adopt iMotions' FACET module, a facial emotion classifier, to extract visual expression features. The acoustic embedding $\mathcal{S}_e^{(3)}$ is extracted by COVAREP \cite{COVAREP} which is a speech processing toolbox. Two-layer stacked LSTM followed by FC layers are used for both visual and acoustic encoders. We leverage a one-layer Transformer encoder in fusion module $\mathcal{C}_f$. The self-attention mechanism is capable of learning the correlation between each modality and extracting collaborative semantic information. An FC layer is set as the prediction module $\mathcal{C}_p$.
\begin{table}[htbp]
    \caption{Specification of distributed multi-modal fusion model}
    \begin{center}
    \begin{threeparttable}
            \begin{tabular}{|m{0.4\textwidth}<{\centering}|m{0.4\textwidth}<{\centering}|}
                \hline
                Module & Implementation  \\
                \hline
                Textual Embedding $\mathcal{S}_e^{(1)}$ & BERT\textcolor{gray}{\tnote{*1}} \\
                \hline
                Visual Embedding $\mathcal{S}_e^{(2)}$   & FACET \\
                \hline
                Acoustic Embedding $\mathcal{S}_e^{(3)}$ & COVAREP \\
                \hline
                Textual Encoder $\mathcal{S}_f^{(1)}$   & FC \\
                \hline
                Visual Encoder $\mathcal{S}_f^{(2)}$     & sLSTM\textcolor{gray}{\tnote{*2}} + FC \\
                \hline
                Acoustic Encoder $\mathcal{S}_f^{(3)}$   & sLSTM + FC \\
                \hline
                Fusion $\mathcal{C}_f$                   & Transformer\textcolor{gray}{\tnote{*3}} \\
                \hline
                Task Function $\mathcal{C}_p$            & FC \\
                \hline
            \end{tabular}
            \begin{tablenotes}
                    \item \textcolor{gray}{*1} \textcolor{gray}{BERT model used here is a 12-layer Transformer.} 
                    \textcolor{gray}{*2} \textcolor{gray}{sLSTM represents the stacked bi-directional LSTM, where the number of layers of sLSTM we use is 2.} 
                    \textcolor{gray}{*3} \textcolor{gray}{We use a one-layer Transformer to fuse three modality features.}
            \end{tablenotes}
            \label{tab:distributed_multimodal_fusion}
        \end{threeparttable}
    \end{center}
    \vspace{-1cm}
\end{table}
\subsubsection{Wireless System Settings}
We implement a distributed multi-modal fusion model in which each edge device is only capable of collecting and processing one type of modality. We assume a flat-fading model, where the channel gain remains static during the transmission of each inference sample. The channel gain for each device is independently generated according to the Rayleigh distribution. In the scenario of additive white Gaussian noise (AWGN) channel, the channel gain satisfies $h^{(m)}=1$. We assume perfect CSI for all edge devices, which is estimated by the server through channel reciprocity in TDD systems. In the simulation, we use low-density parity-check (LDPC) code for channel coding. The channel coding is based on 3GPP specification \cite{3GPP-38.214}. The physical layer supports modulations of QPSK, $16$-QAM, $64$-QAM, and $256$-QAM. For simplicity, we assume that all edge devices transmit with the same power $P^{(m)}=P,\ \forall m\in\{1,2,\cdots,M\}$. 
We set the bound of semantic task distortion to $\Delta_0 = 10^{-3}$, as it balances the need to avoid severe degradation of inference accuracy resulting from a too large bound against the need to prevent excessive communication resources required for each modality from too small bounds.

\subsubsection{Baselines}

Regarding the baseline schemes in the simulation, we adopt both the conventional transmission approach and the NN-based JSCC approach in semantic communications.
\begin{itemize}
    \item Conventional method: This method uses the same NN as the source encoder, followed by conventional physical layer transmission. In this approach, we also use LDPC codes for channel coding but the code rate for all modalities are the same, i.e., $R^{(m)}=R^{\text{fixed}},\ \forall m\in\{1,2,\cdots,M\}$. To make a fair comparison, we assume the same transmission delay in both conventional methods and our proposed mechanism. Therefore, the rate adopted in this approach is determined by
    \begin{equation}
        R^{\text{fixed}} = \max_m \frac{D^{(m)}}{\max_m\frac{D^{(m)}}{R^{(m)}}}.
    \end{equation}
    \item JSCC: To perform an end-to-end NN-based coding system, we modify the JSCC scheme in \cite{DeepSC}. We add a layer simulating a wireless channel in the original multi-modal fusion network between semantic encoders $\mathcal{S}_{\text{JSCC}}^{(m)}$ and semantic decoder $\mathcal{C}_{\text{JSCC}}$. In this approach, we set the channel layer as the AWGN channel with $\text{SNR}=0\text{dB}$. Using the aforementioned pre-trained MISA model, we freeze the semantic encoders and only fine-tune the semantic decoder based on the loss function given in \cite{DeepSC}. Then, the model $\mathcal{F}^{\text{JSCC}}$ is separated and deployed at edge devices and the server respectively. Note that JSCC views the output tensor of the NN as constellation points. Considering the feasibility in communication systems, we quantize the double precision signals into the $4096$-QAM constellation, which is large enough to mimic the continuous constellation in \cite{JSCC-image} and \cite{DeepSC}.
\end{itemize}
\begin{table}[htb]
    \caption{Performance of distributed multi-modal fusion through error-free transmission}
    \begin{center}
        \begin{tabular}{|m{0.2\textwidth}<{\leftraggle}|m{0.2\textwidth}<{\centering}||m{0.2\textwidth}<{\centering}|m{0.2\textwidth}<{\centering}|m{0.2\textwidth}<{\centering}|}
            \hline
            \multicolumn{1}{|p{0.2\textwidth}<{\centering}}{Models}                 & \multicolumn{2}{|p{0.3\textwidth}<{\centering}}{MOSI}                   & \multicolumn{2}{|p{0.3\textwidth}<{\centering}|}{MOSEI}                 \\
            \hline
            \multicolumn{1}{|p{0.2\textwidth}<{\centering}}{Metrics}                & \multicolumn{1}{|p{0.15\textwidth}<{\centering}}{MAE ($\downarrow$)}    & \multicolumn{1}{|p{0.15\textwidth}<{\centering}}{Corr ($\uparrow$)}     & \multicolumn{1}{|p{0.15\textwidth}<{\centering}}{MAE ($\downarrow$)}    & \multicolumn{1}{|p{0.15\textwidth}<{\centering}|}{Corr ($\uparrow$)}    \\
            \hline
            \multicolumn{1}{|p{0.2\textwidth}<{\centering}}{MulT}                   & \multicolumn{1}{|p{0.15\textwidth}<{\centering}}{0.871}                 & \multicolumn{1}{|p{0.15\textwidth}<{\centering}}{0.698}                 & \multicolumn{1}{|p{0.15\textwidth}<{\centering}}{0.580}                 & \multicolumn{1}{|p{0.15\textwidth}<{\centering}|}{0.700}                \\
            \hline
            \multicolumn{1}{|p{0.2\textwidth}<{\centering}}{ICCN}                   & \multicolumn{1}{|p{0.15\textwidth}<{\centering}}{0.860}                 & \multicolumn{1}{|p{0.15\textwidth}<{\centering}}{0.710}                 & \multicolumn{1}{|p{0.15\textwidth}<{\centering}}{0.565}                 & \multicolumn{1}{|p{0.15\textwidth}<{\centering}|}{0.713}                \\
            \hline
            \multicolumn{1}{|p{0.2\textwidth}<{\centering}}{MISA}                   & \multicolumn{1}{|p{0.15\textwidth}<{\centering}}{0.783}                 & \multicolumn{1}{|p{0.15\textwidth}<{\centering}}{0.761}                 & \multicolumn{1}{|p{0.15\textwidth}<{\centering}}{0.555}                 & \multicolumn{1}{|p{0.15\textwidth}<{\centering}|}{0.756}                \\
            \hline
            \multicolumn{1}{|p{0.2\textwidth}<{\centering}}{Ours}                   & \multicolumn{1}{|p{0.15\textwidth}<{\centering}}{0.784}                 & \multicolumn{1}{|p{0.15\textwidth}<{\centering}}{0.761}                 & \multicolumn{1}{|p{0.15\textwidth}<{\centering}}{0.554}                 & \multicolumn{1}{|p{0.15\textwidth}<{\centering}|}{0.752}                \\
            \hline
        \end{tabular}
        \label{tab:multimodal_fusion_performance}
    \end{center}
    \vspace{-1cm}
\end{table}
\subsection{Performance Analysis}
\label{sec:performance_analysis}
\subsubsection{Error-Free Performance}
First, we evaluate the performance of the distributed multi-modal fusion using error-free transmission to demonstrate the effectiveness of the multi-modal fusion model. Additionally, the error-free transmission performance could serve as the upper bound for simulations in wireless channels, since the full modal semantic features $U^{(m)}$ are transmitted to the server noiselessly. We choose three existing multi-modal fusion models for the task of MSA: MulT \cite{MulT} which utilizes the self-attention mechanism to learn cross-modality representations, ICCN \cite{ICCN} which uses the outer product between modalities for canonical correlation analysis, and MISA \cite{MISA} which combines modality invariant and specific representations to analyze both characteristic and private features. 
% Note that, MISA also stands for the state-of-art (SOTA) model. 

\begin{figure}[htbp]
    \centering
    \subfigure[MOSI dataset.]{
        \begin{minipage}[t]{0.45\textwidth}
        \centering
        \includegraphics[width=\textwidth]{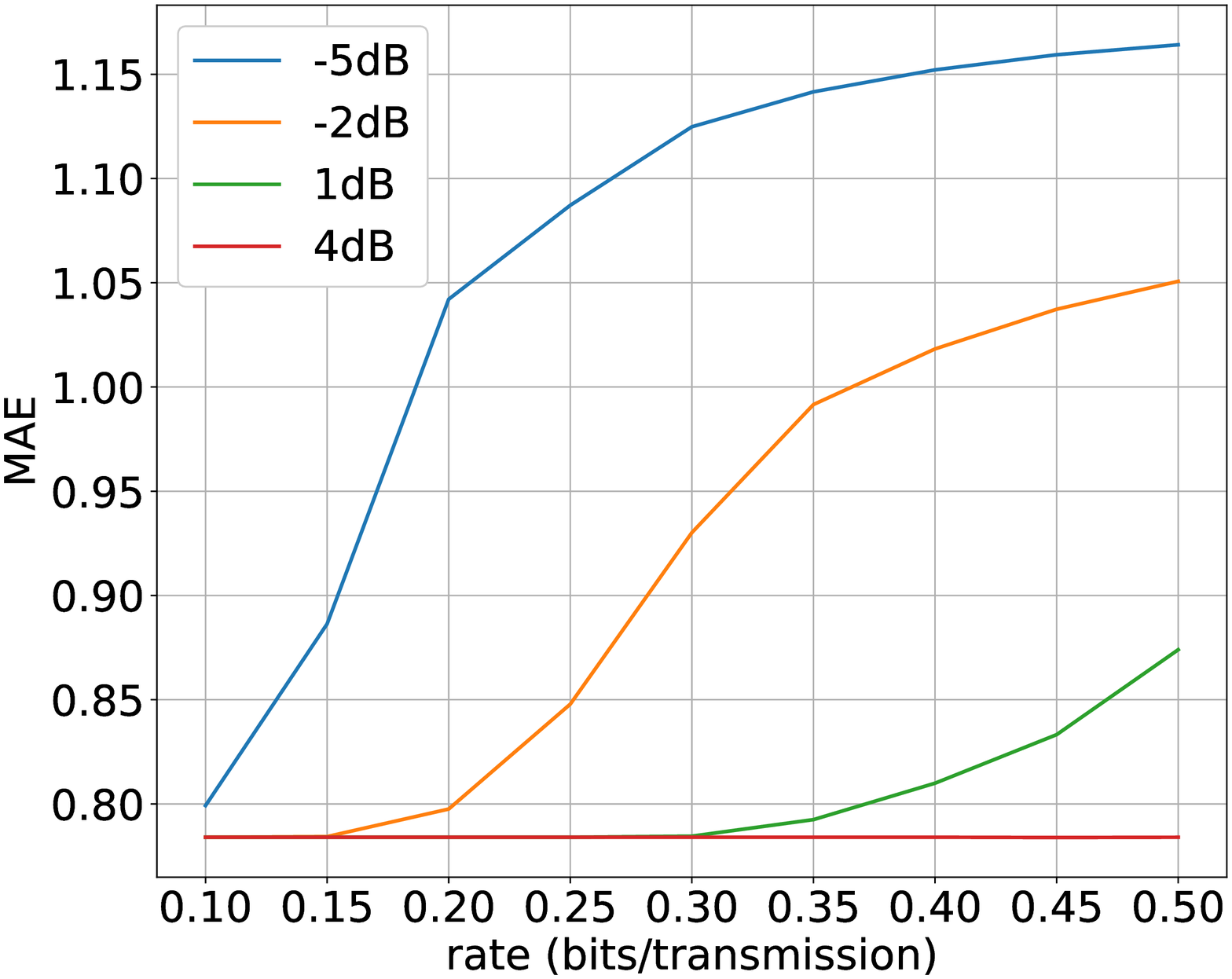}
        \end{minipage}
        \begin{minipage}[t]{0.45\textwidth}
        \centering
        \includegraphics[width=\textwidth]{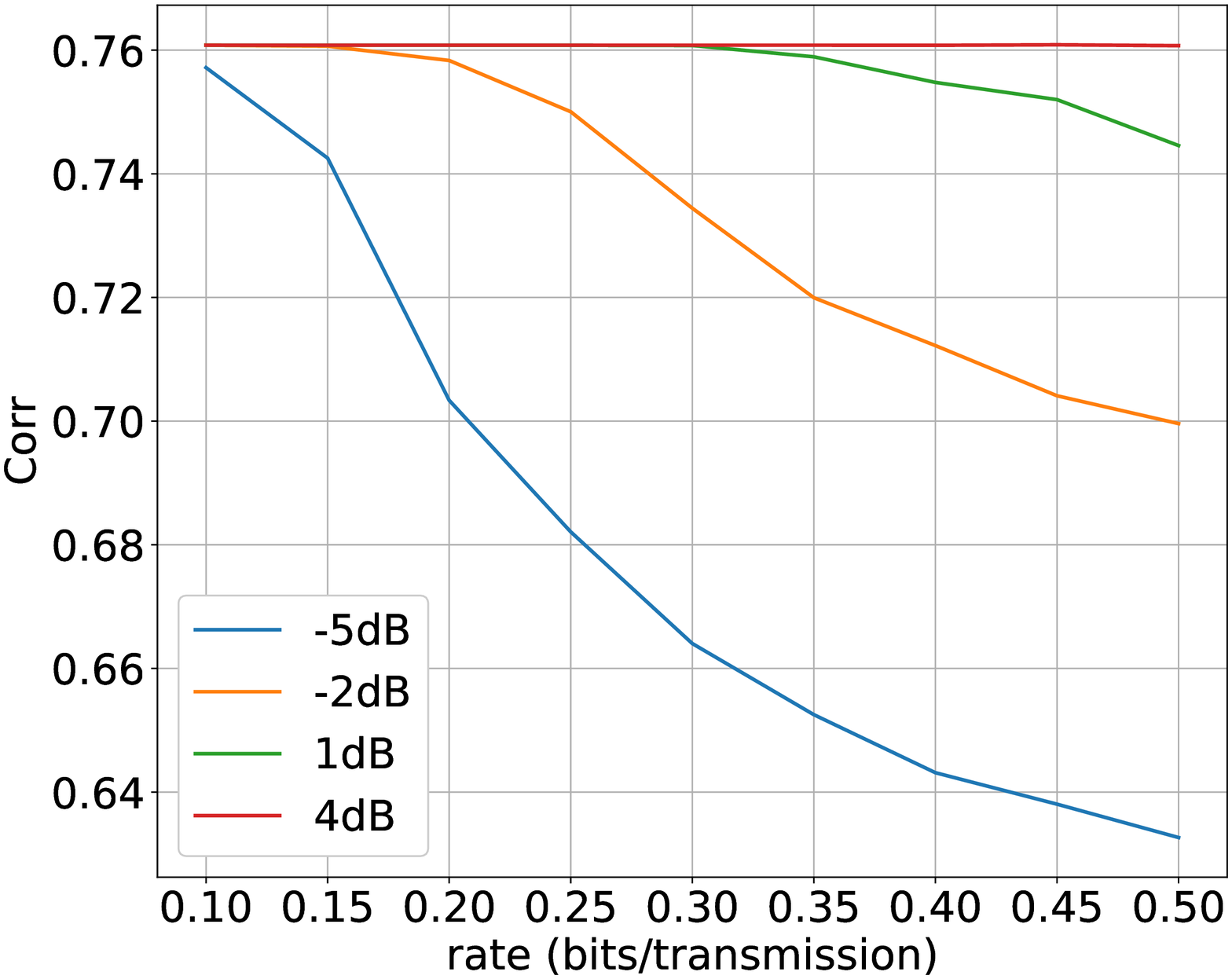}
        \end{minipage}
    }
    
    \subfigure[MOSEI dataset.]{
        \begin{minipage}[t]{0.45\textwidth}
        \centering
        \includegraphics[width=\textwidth]{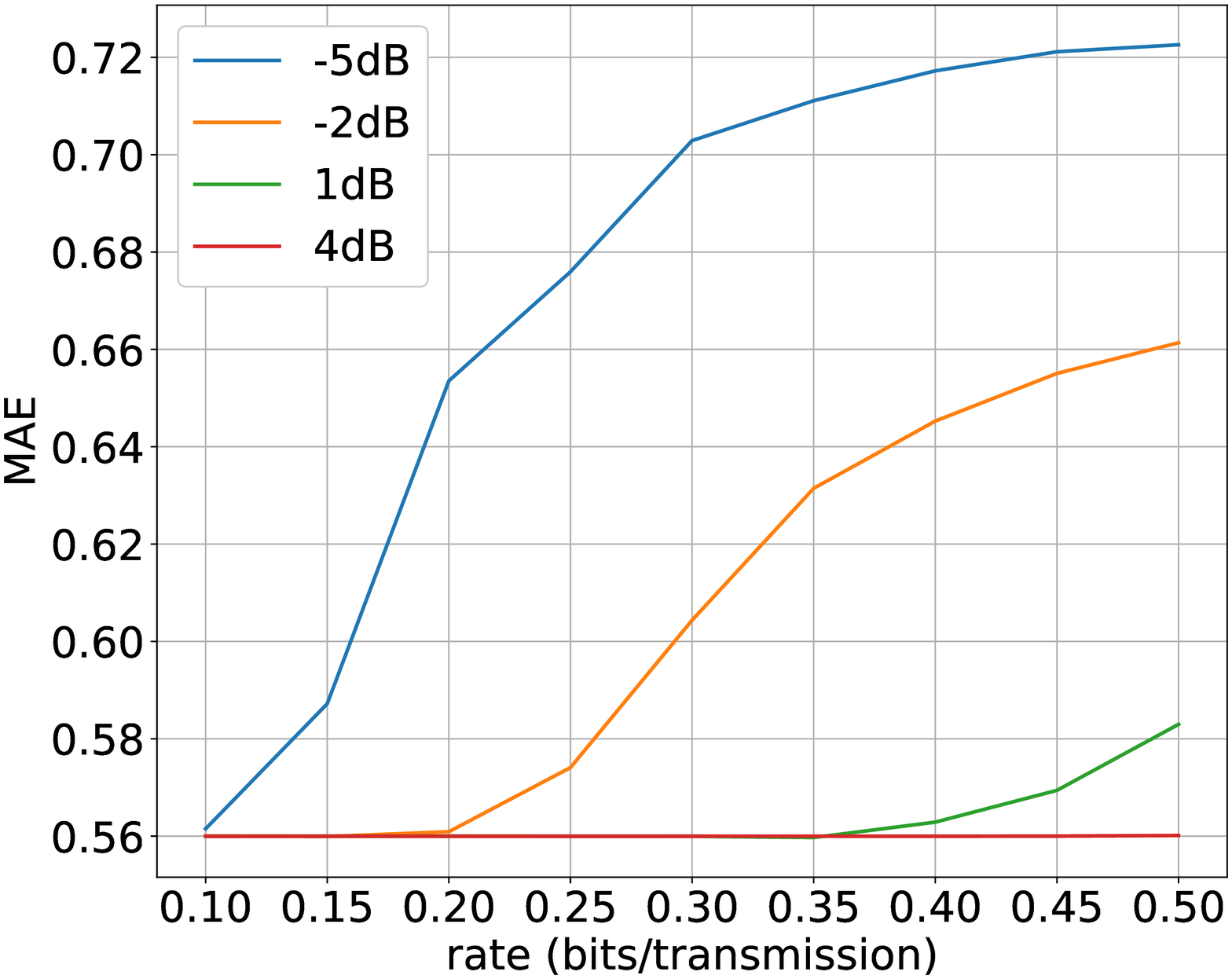}
        \end{minipage}
        \begin{minipage}[t]{0.45\textwidth}
        \centering
        \includegraphics[width=\textwidth]{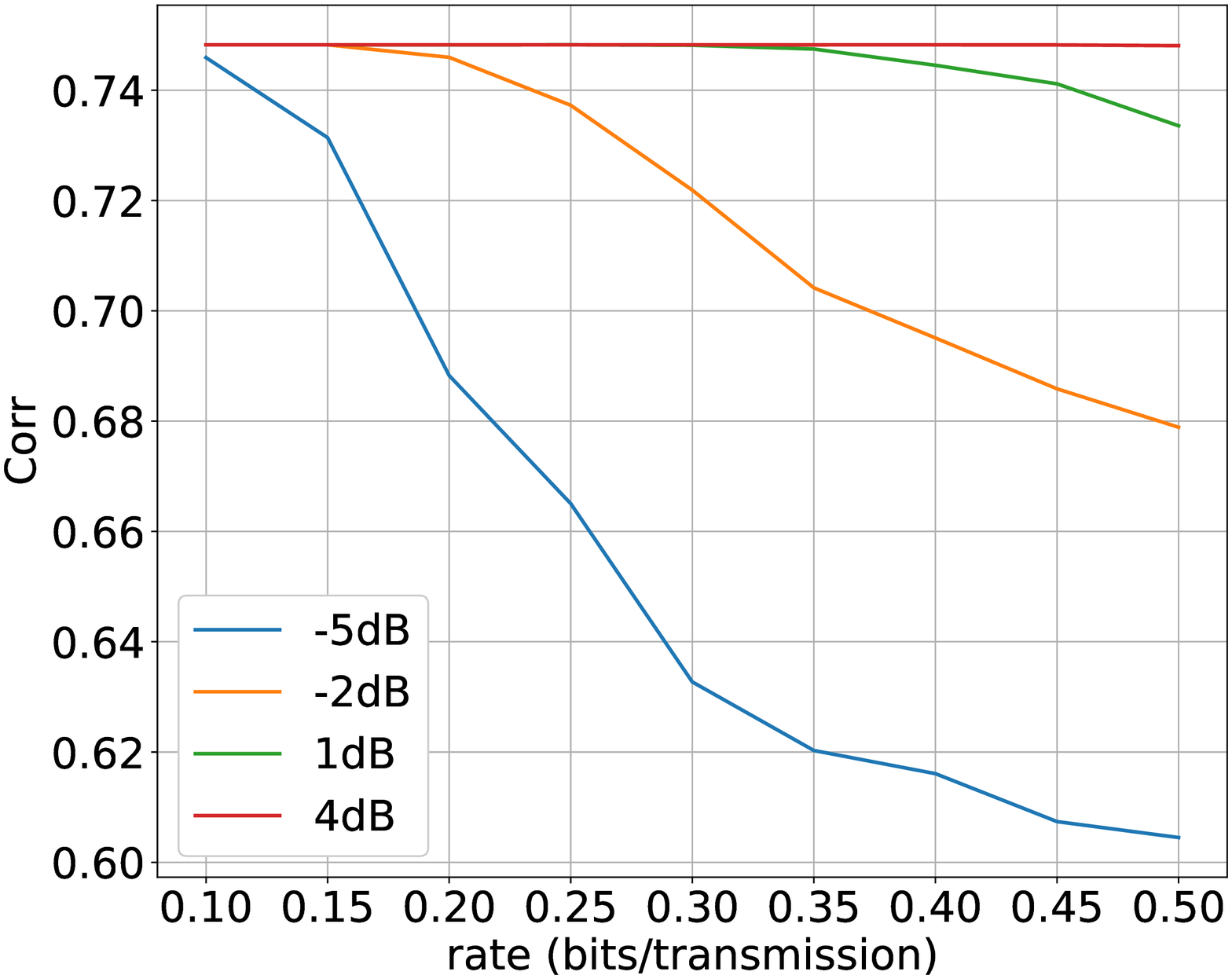}
        \end{minipage}
    }
    \caption{Performance of distributed multi-modal fusion model with the same coding rate for all modalities in the AWGN channel.}
    \label{fig:awgn_snr_rate}
\end{figure}

\begin{figure}[htbp]
    \centering
    \subfigure[MOSI dataset.]{
        \begin{minipage}[t]{0.45\textwidth}
        \centering
        \includegraphics[width=\textwidth]{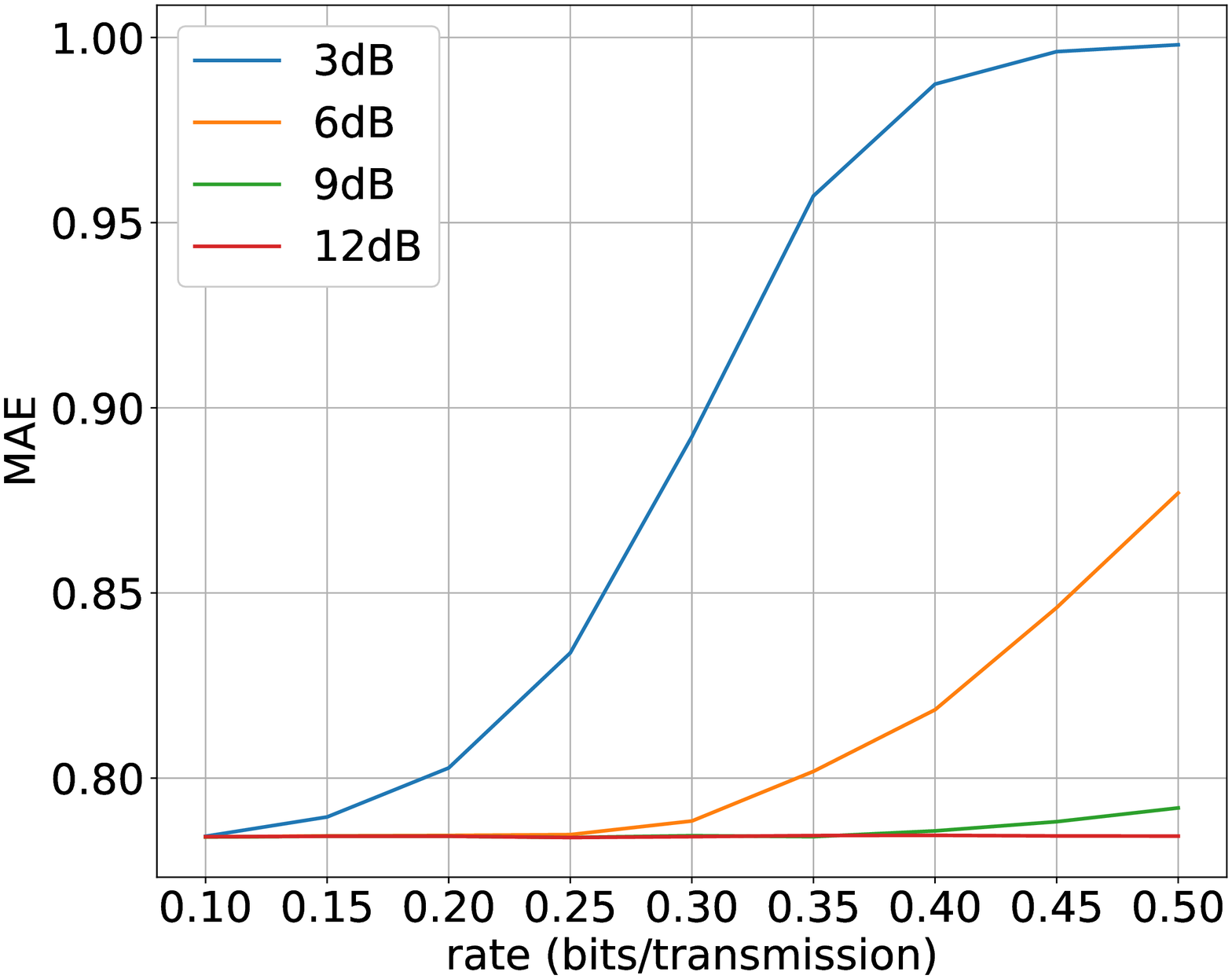}
        \end{minipage}
        \begin{minipage}[t]{0.45\textwidth}
        \centering
        \includegraphics[width=\textwidth]{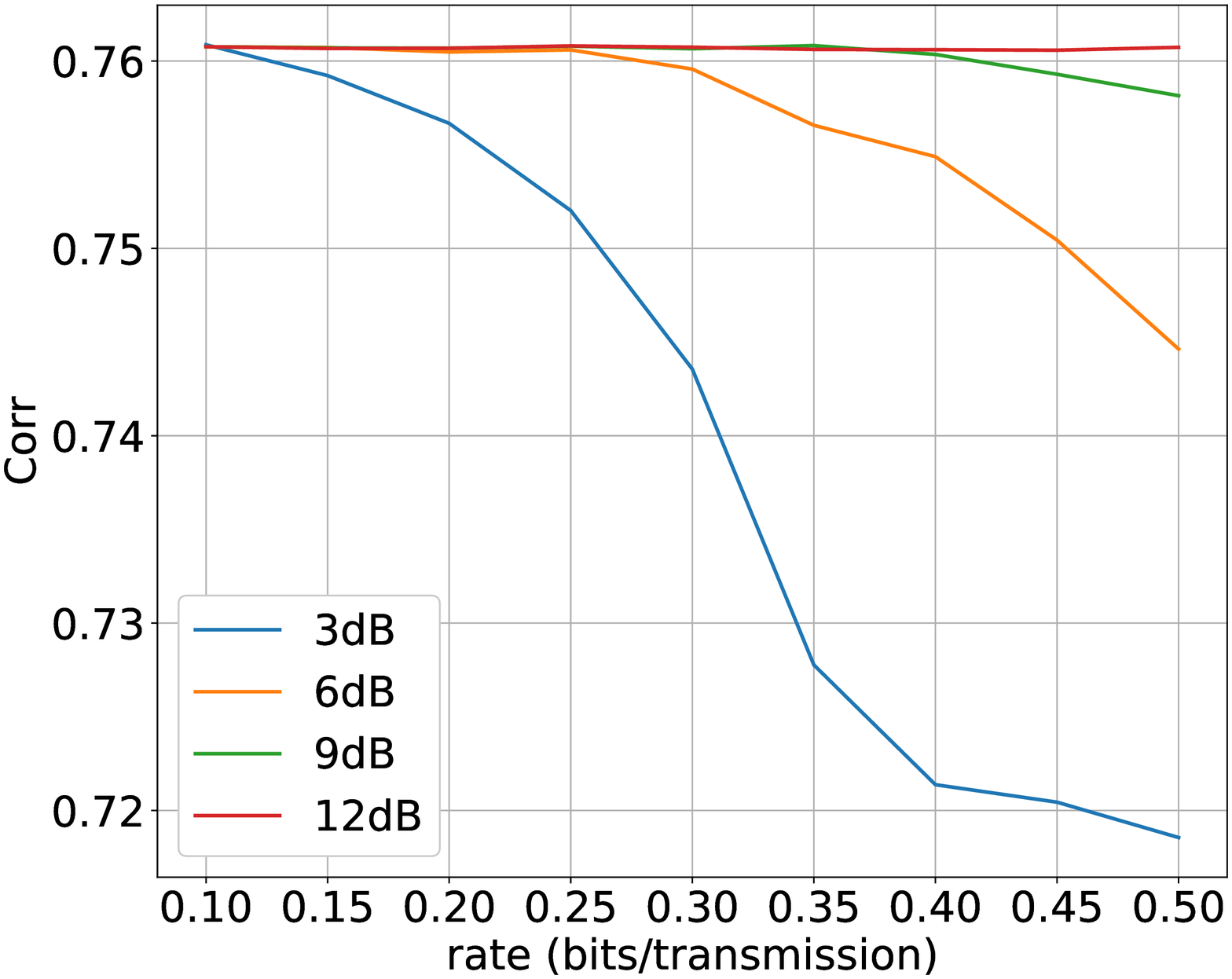}
        \end{minipage}
    }
    
    \subfigure[MOSEI dataset.]{
        \begin{minipage}[t]{0.45\textwidth}
        \centering
        \includegraphics[width=\textwidth]{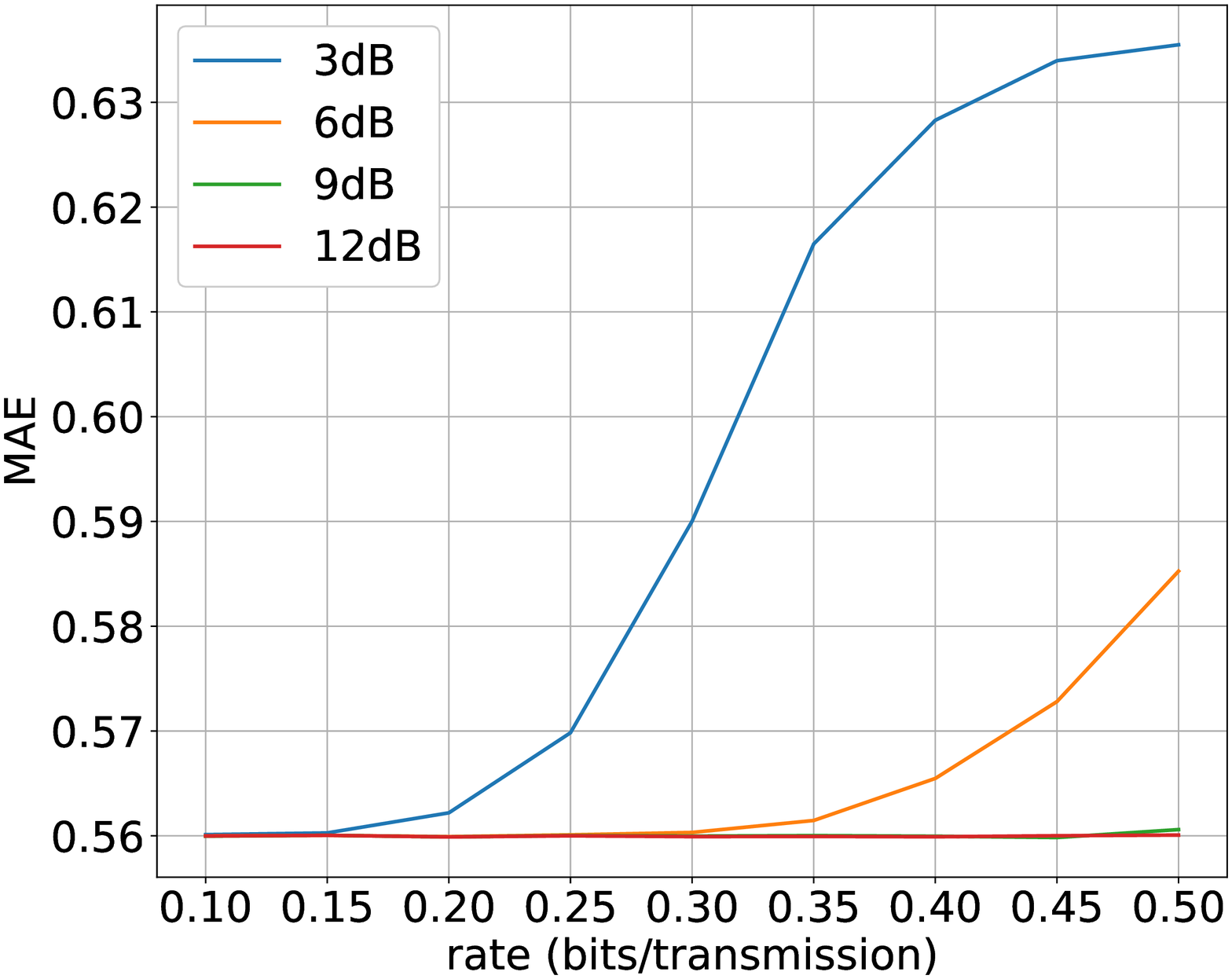}
        \end{minipage}
        \begin{minipage}[t]{0.45\textwidth}
        \centering
        \includegraphics[width=\textwidth]{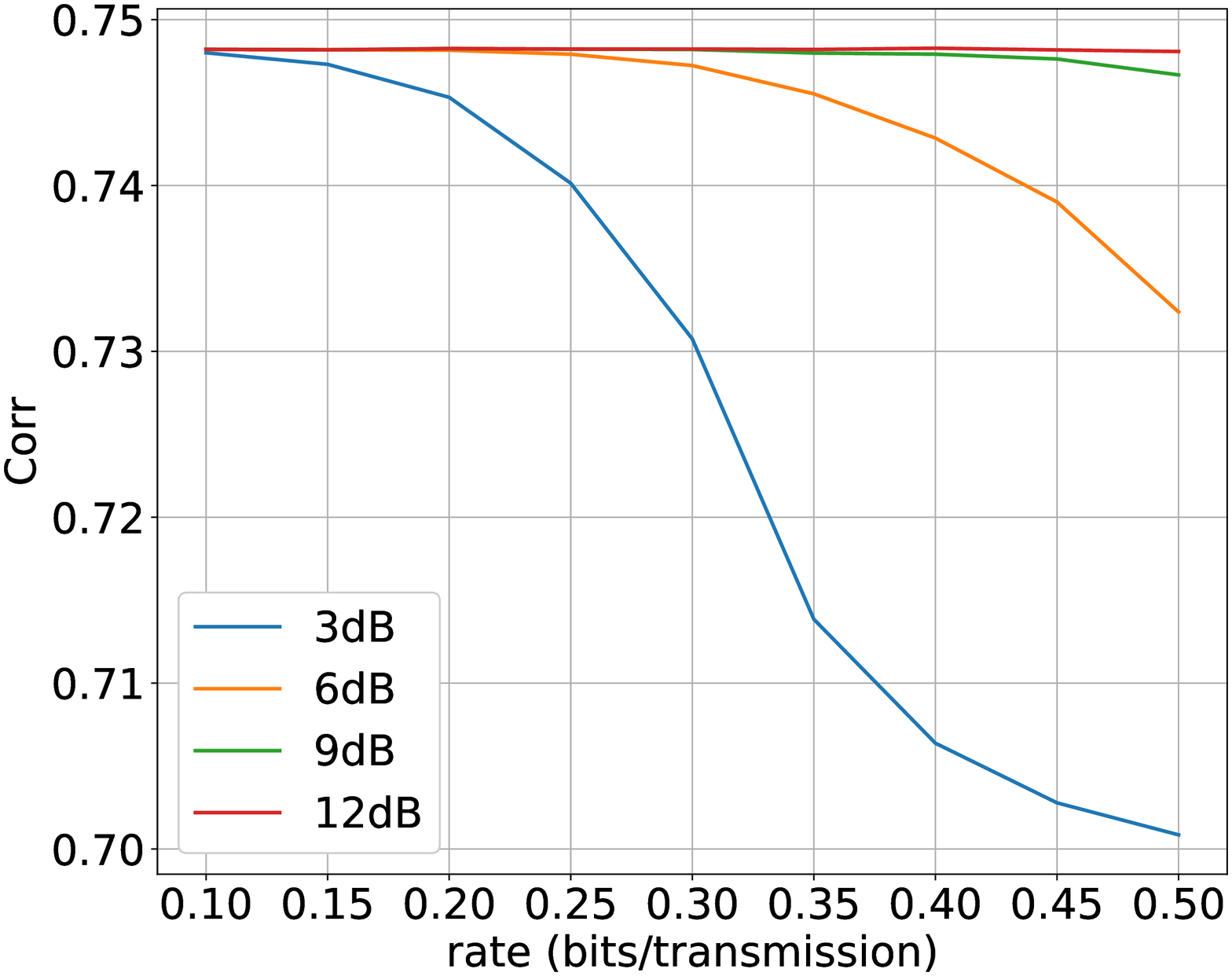}
        \end{minipage}
    }
    \caption{Performance of distributed multi-modal fusion model with the same coding rate for all modalities in the Rayleigh channel.}
    \label{fig:rayleigh_snr_rate}
    \vspace{-0.5cm}
\end{figure}

The comparative results on the MOSI and MOSEI datasets are presented in \autoref{tab:multimodal_fusion_performance}. From the results, the proposed distributed multi-modal fusion model achieves almost the same performance as MISA. To separating the NN into a semantic encoder and a semantic decoder in our distributed multi-modal semantic communication framework, we discard some intricate information extraction methods and modality fusion techniques in the original model without sacrificing the overall inference performance. Even without sophisticated fusion methods, the proposed model fares better than most baselines and maintains a neglectable gap with SOTA.

\subsubsection{The Impact of Coding Rate}

To illustrate the importance of coding rate selection in the distributed multi-modal semantic communication framework, we test the conventional method over multiple code rates in both AWGN and Rayleigh channels, and the performance results are shown in \autoref{fig:awgn_snr_rate} and \autoref{fig:rayleigh_snr_rate}. From the figures, reducing the channel coding rate is an effective means to fulfill the requirement of semantic tasks. Increasing the transmission rate to reduce inference delay inevitably impairs the accuracy of semantic tasks in the low SNR scenario. 

Furthermore, the figures demonstrate that using an equal transmission rate for all modalities makes performance sensitive to the channel SNR. This result arises because the variability in semantic importance among modalities plays a significant role in semantic tasks.
Generally, both task performance and communication efficiency are limited by the modality that is with a large data size but less semantically crucial. Therefore, it is of great importance to assign unequal error protection to different modalities to balance the semantic task performance and transmission efficiency. Specifically, while increasing the transmission rate helps reduce delay for modalities with large data sizes, it is important to maintain a low rate for semantically significant modalities across various SNRs to ensure inference accuracy.
% \autoref{fig:awgn_snr_rate} and \autoref{fig:rayleigh_snr_rate} also demonstrate that the optimal rate for the crucial modality under different SNRs corresponds to the cross-point between the current SNR and the highest SNR. For example, in \autoref{fig:awgn_snr_rate} (b), we can observe that for the MOSEI dataset under the AWGN channel, the optimal rates for the most important modality across different SNRs are around 0.1 at -5dB, 0.2 at -2dB, and 0.35 at 1dB.} 

% \begin{figure}[hb]
%     \centering
%     \subfigure[MOSI dataset.]{
%         \begin{minipage}[t]{0.45\textwidth}
%         \centering
%         \includegraphics[width=\textwidth]{figures/problem2_mosi.eps}
%         \end{minipage}
%     }
%     \subfigure[MOSEI dataset.]{
%         \begin{minipage}[t]{0.45\textwidth}
%         \centering
%         \includegraphics[width=\textwidth]{figures/problem2_mosei.eps}
%         \end{minipage}
%     }
%     \caption{Optimal transmission rates for textual, visual, and acoustic modality.}
%     \label{fig:optimal_rate}
% \end{figure}

\subsubsection{Optimal Rates}

\begin{figure}[htbp]
    \centering
    \subfigure[MOSI dataset.]{
        \begin{minipage}[t]{0.45\textwidth}
        \centering
        \includegraphics[width=\textwidth]{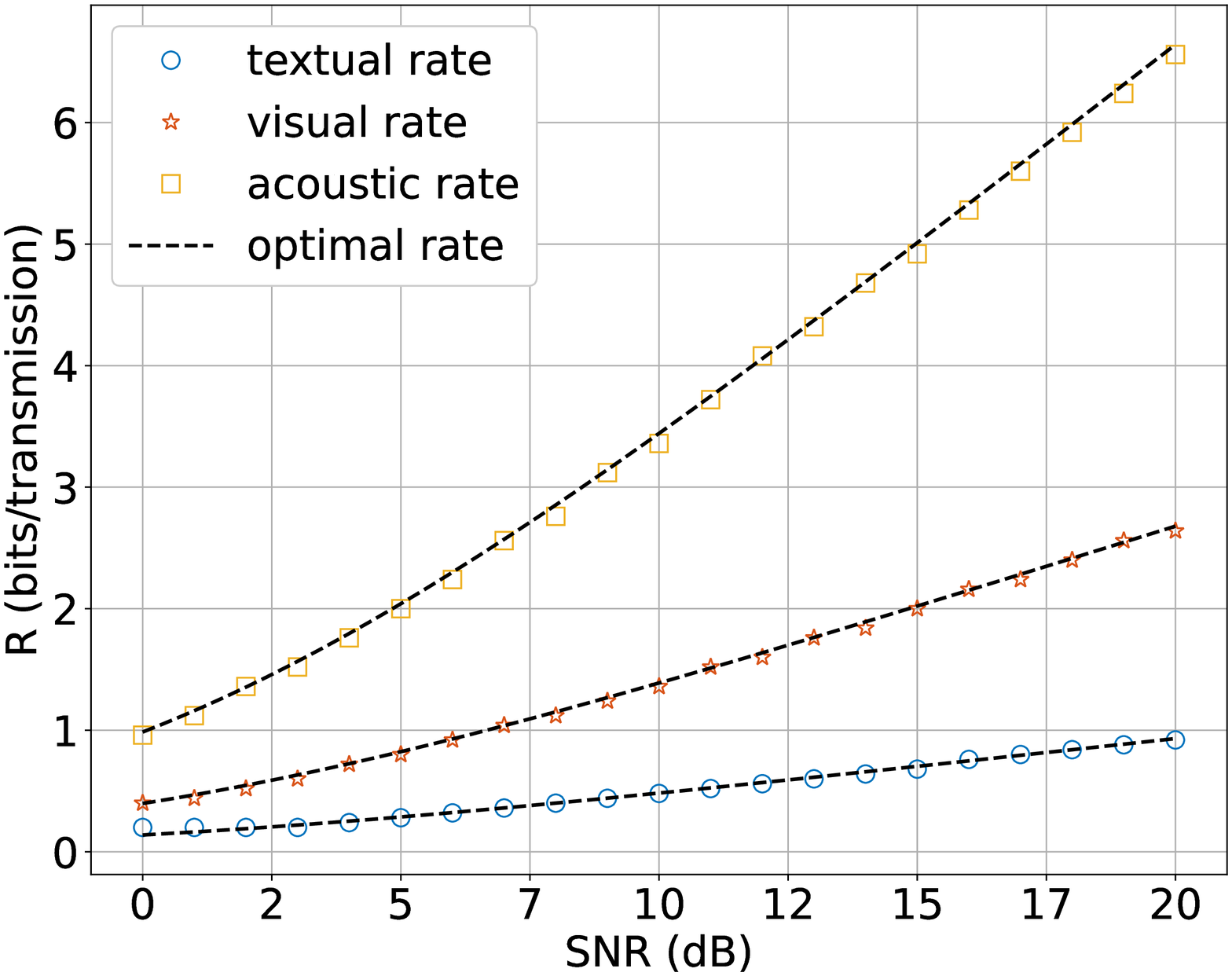}
        \end{minipage}
    }
    \subfigure[MOSEI dataset.]{
        \begin{minipage}[t]{0.45\textwidth}
        \centering
        \includegraphics[width=\textwidth]{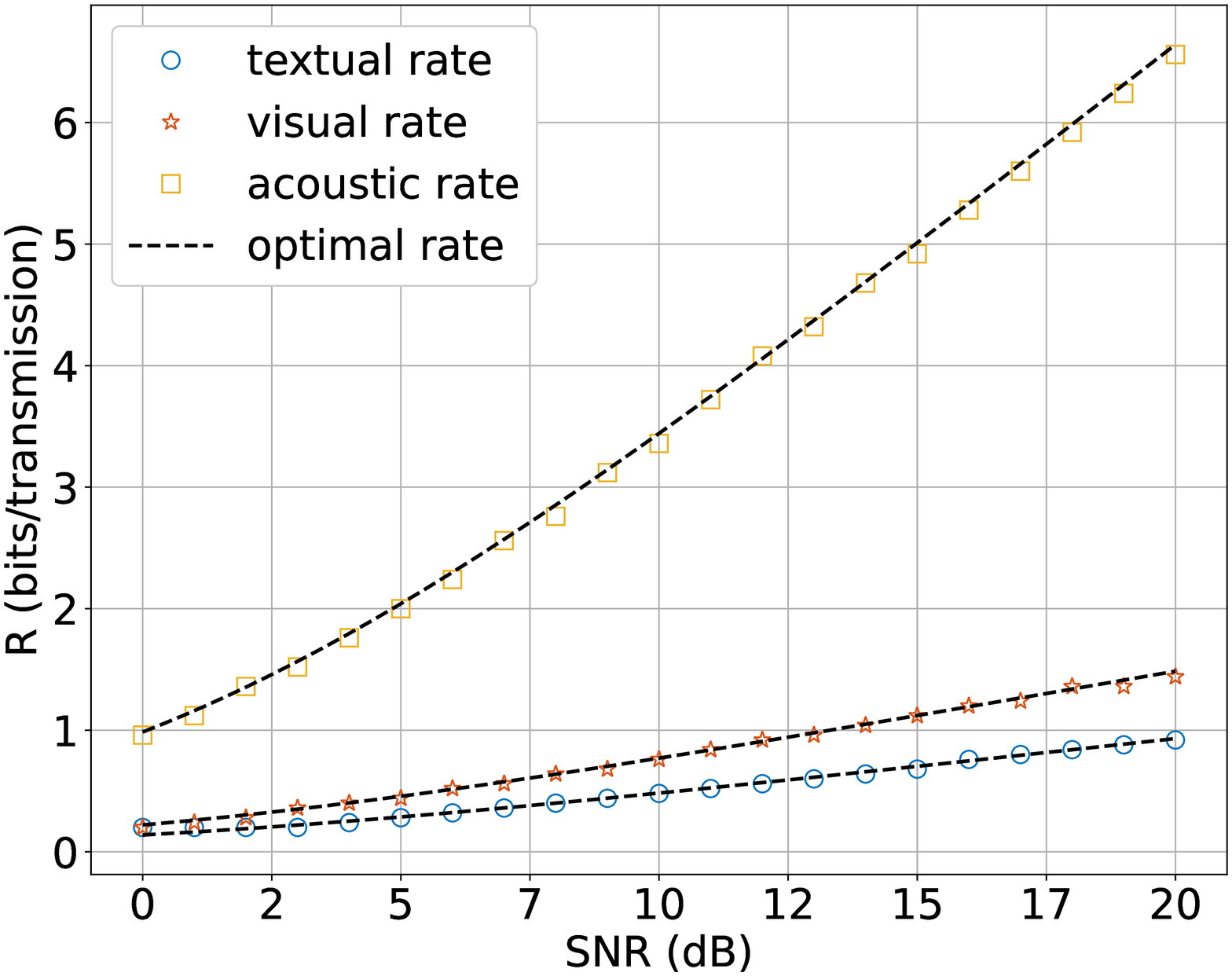}
        \end{minipage}
    }
    \caption{Actual and optimal transmission rates for textual, visual, and acoustic modality in distributed multi-modal semantic communication framework.}
    \label{fig:actual_rate}
\end{figure}

To demonstrate the effectiveness of jointly considering semantic encoders and decoders during transmission, we plot the optimal rate by solving $\mathcal{P}2$ in \autoref{fig:actual_rate}. In the distributed multi-modal semantic communication framework, modality features are encoded at a rate based on the solution of optimization. Each edge device determines the rate of the LDPC code and the modulation order that approximate the optimal transmission rate. The actual transmission rates adopted for three modalities are also illustrated in \autoref{fig:actual_rate}. With the feature of rate matching in the 3GPP standard, the system supports a more flexible rate selection. As a result, the performance loss due to the quantization of the optimal solution can be reasonably neglected in the practical system.

From the figures, the transmission rates for all modalities increase with the channel SNR.
The optimal rate for each modality varies based on both the semantic importance and the data size. 
The size of the textual modality is relatively small and it is semantically important to the sentiment analysis task. 
As a result, the optimal rate for textual modality is the lowest among all modalities. Transmitting at a low coding rate significantly reduces the error probability and achieves better semantic task performance. 
Meanwhile, even though the transmission delay of textual modality increases, it will not affect the end-to-end inference delay because of the small amount of data. 
On the contrary, transmitting visual and acoustic modality would lead to a large communication overhead. Thus, the optimal rate for these modalities becomes higher to keep a small communication delay. Additionally, the acoustic data contains less semantic information and plays a trivial role in the MSA task. 
As shown in \autoref{fig:actual_rate}, the optimal acoustic rate is considerably larger than other rates, which indicates that the semantic task allows more distortion for acoustic features.

\subsubsection{Performance Comparisons}

\begin{figure}[htbp]
    \centering
        \subfigure[MOSI dataset.]{
            \begin{minipage}[t]{0.45\textwidth}
            \centering
            \includegraphics[width=\textwidth]{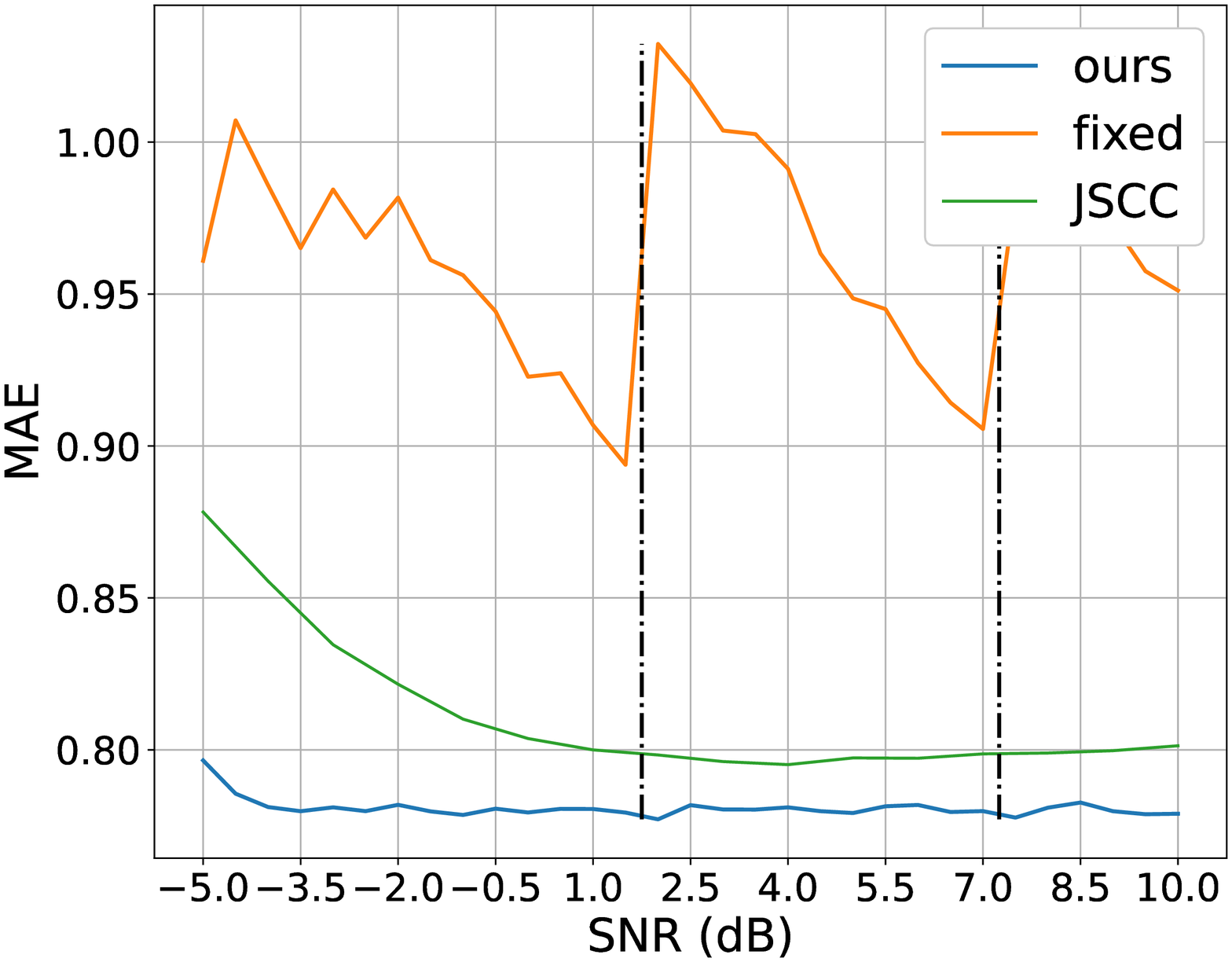}
            \end{minipage}
            \begin{minipage}[t]{0.45\textwidth}
            \centering
            \includegraphics[width=\textwidth]{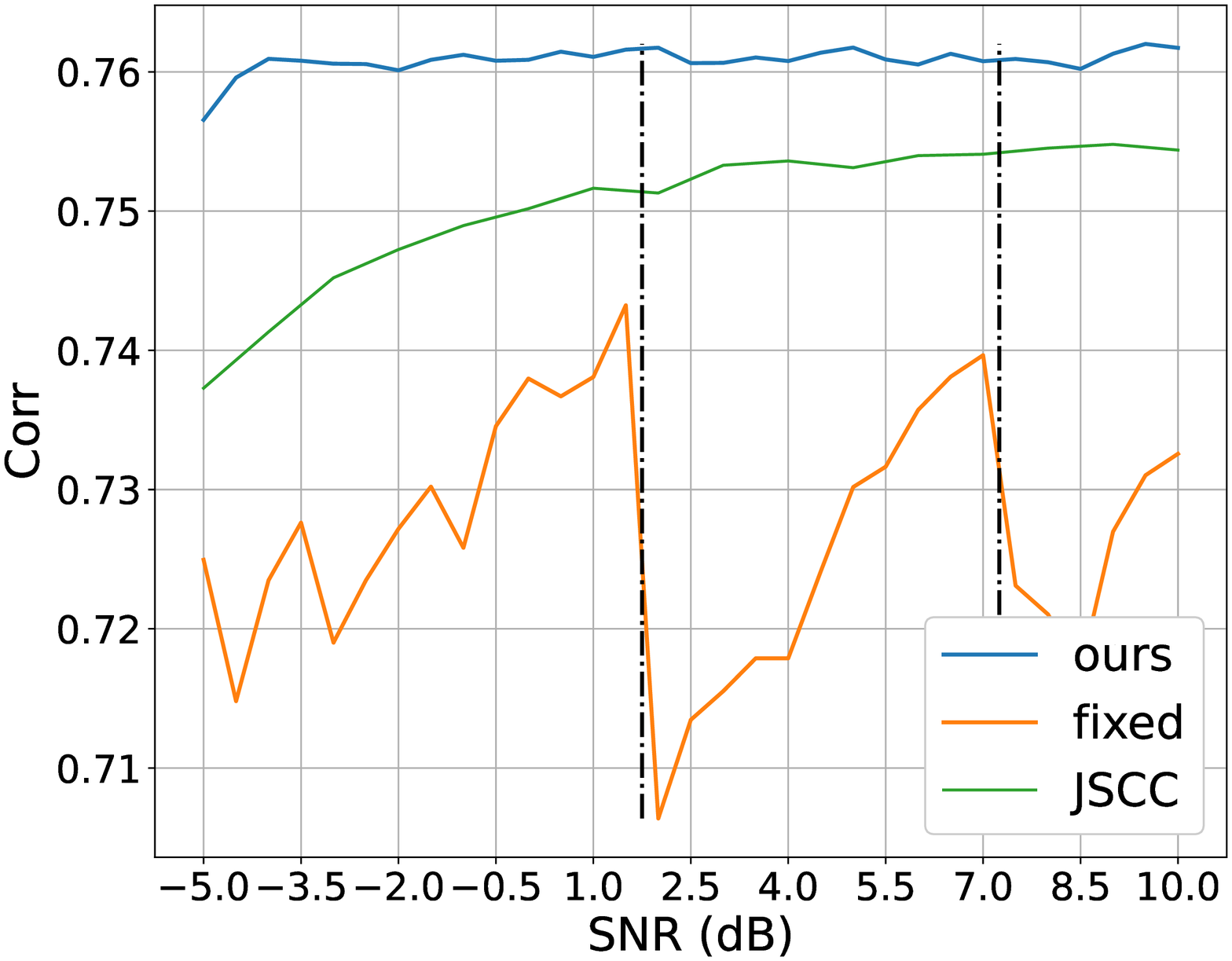}
            \end{minipage}
        }
        
        \subfigure[MOSEI dataset.]{
            \begin{minipage}[t]{0.45\textwidth}
            \centering
            \includegraphics[width=\textwidth]{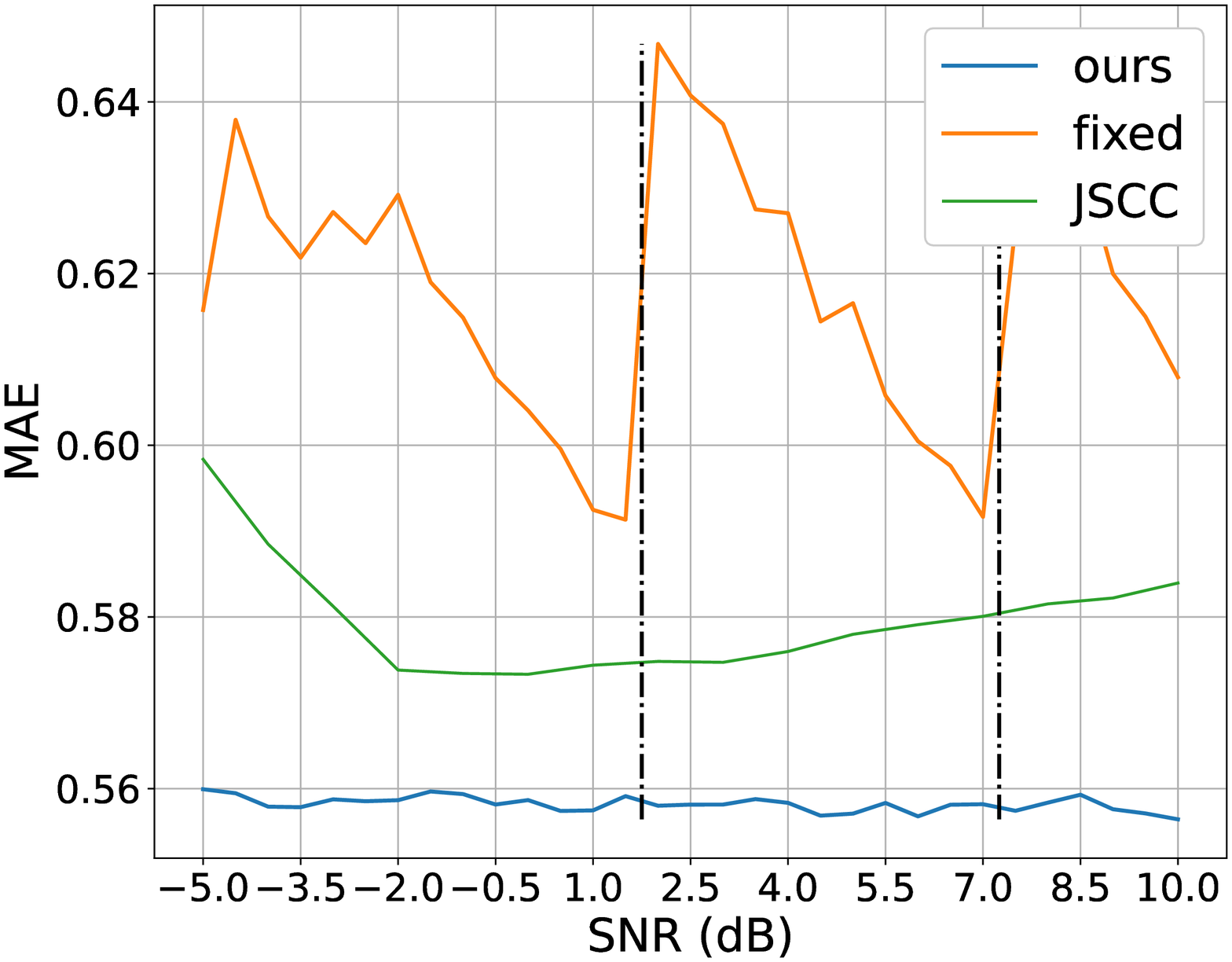}
            \end{minipage}
            \begin{minipage}[t]{0.45\textwidth}
            \centering
            \includegraphics[width=\textwidth]{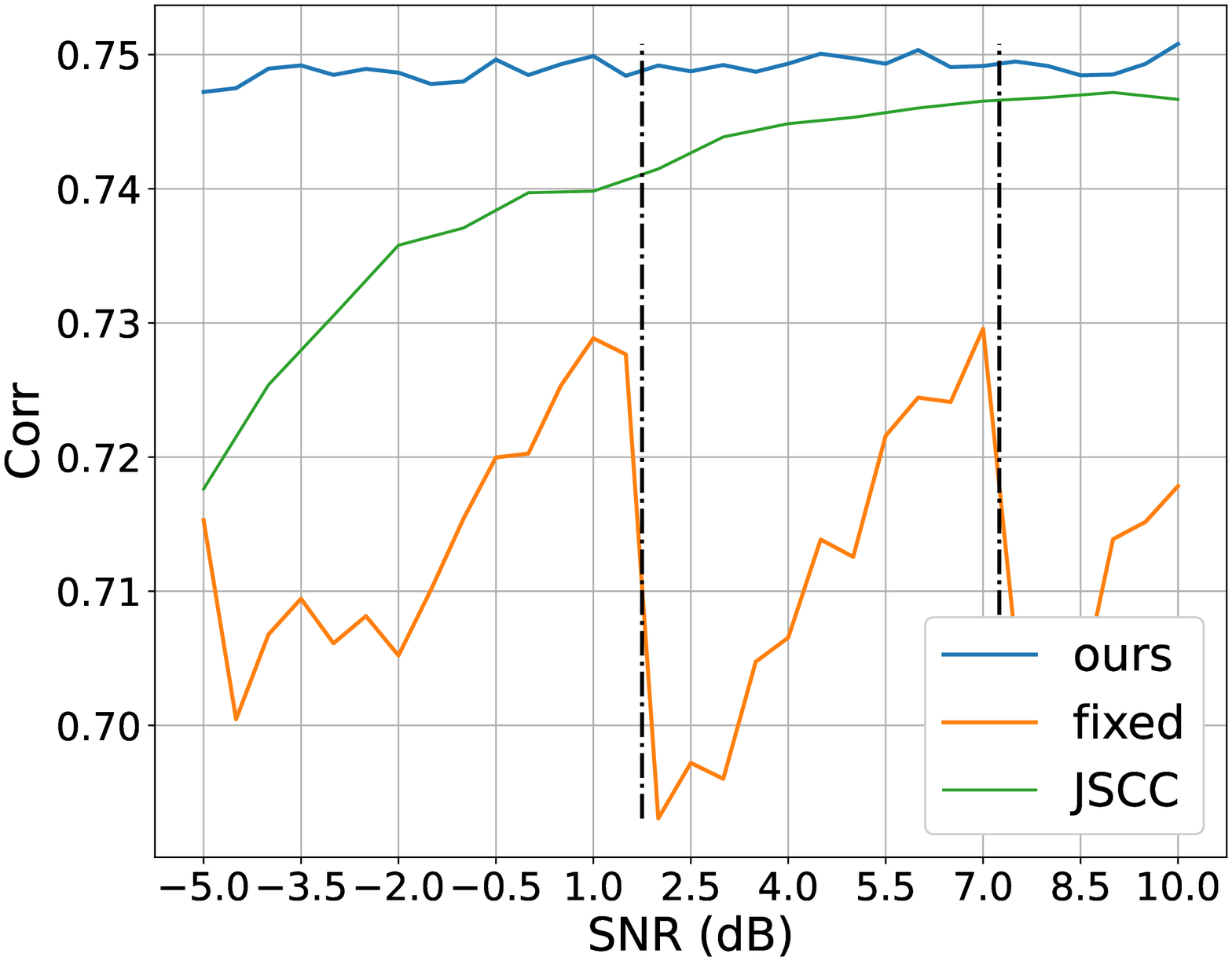}
            \end{minipage}
        }
    \caption{Performance comparison in the AWGN channel.}
    \label{fig:comparison_AWGN}
    \vspace{-0.5cm}
\end{figure}

\begin{figure}[htbp]
    \centering
        \subfigure[MOSI dataset.]{
            \begin{minipage}[t]{0.45\textwidth}
            \centering
            \includegraphics[width=\textwidth]{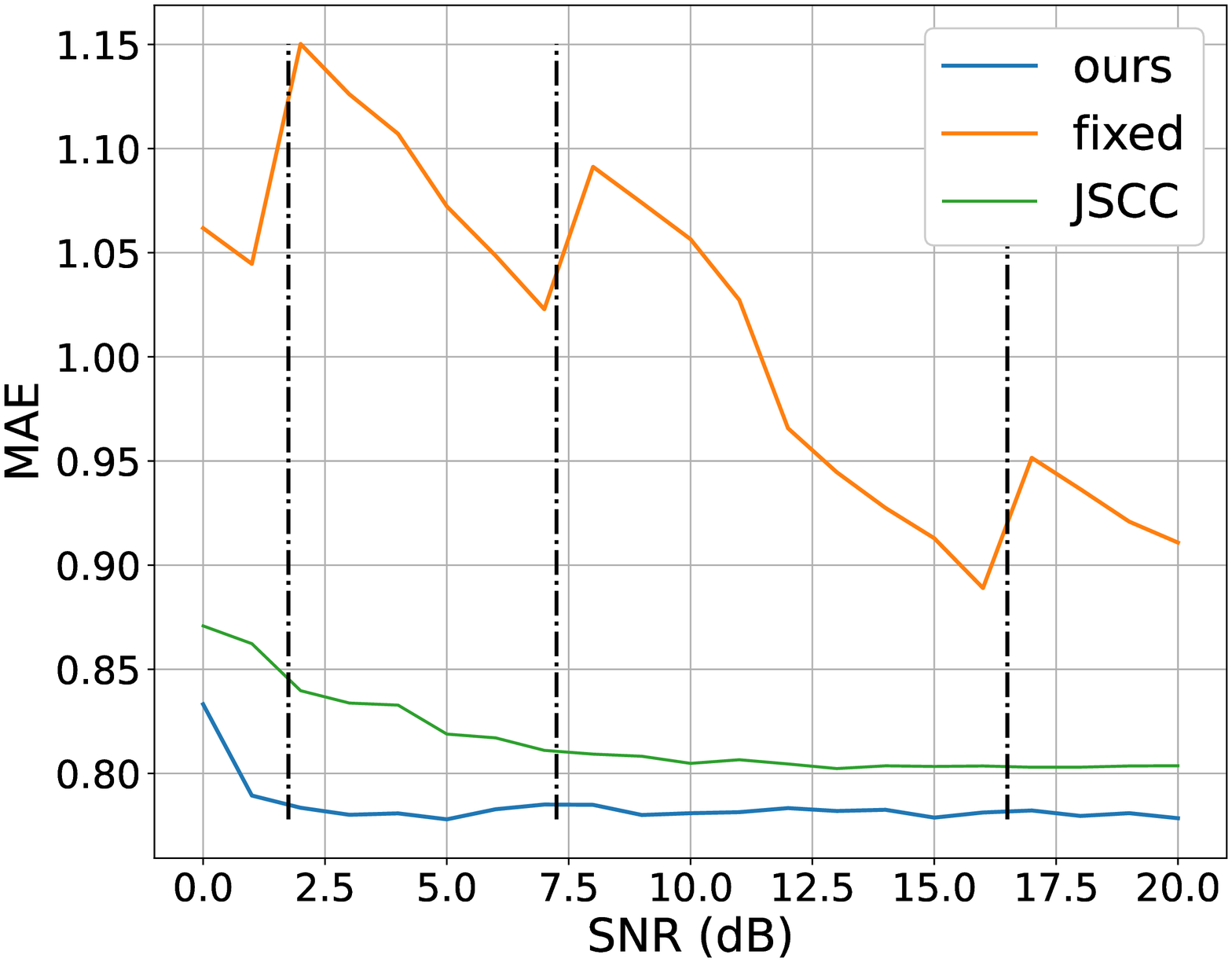}
            \end{minipage}
            \begin{minipage}[t]{0.45\textwidth}
            \centering
            \includegraphics[width=\textwidth]{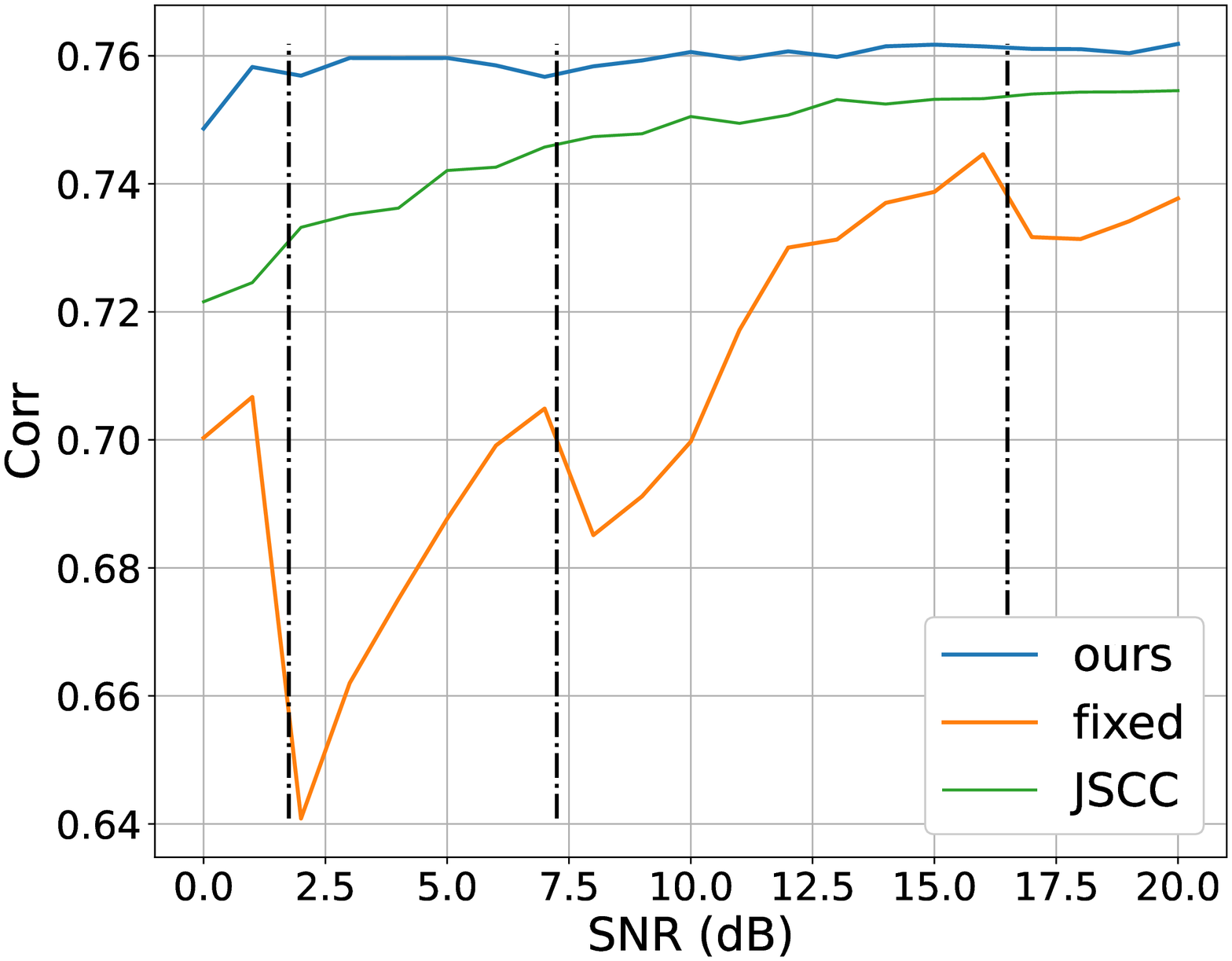}
            \end{minipage}
        }
        
        \subfigure[MOSEI dataset.]{
            \begin{minipage}[t]{0.45\textwidth}
            \centering
            \includegraphics[width=\textwidth]{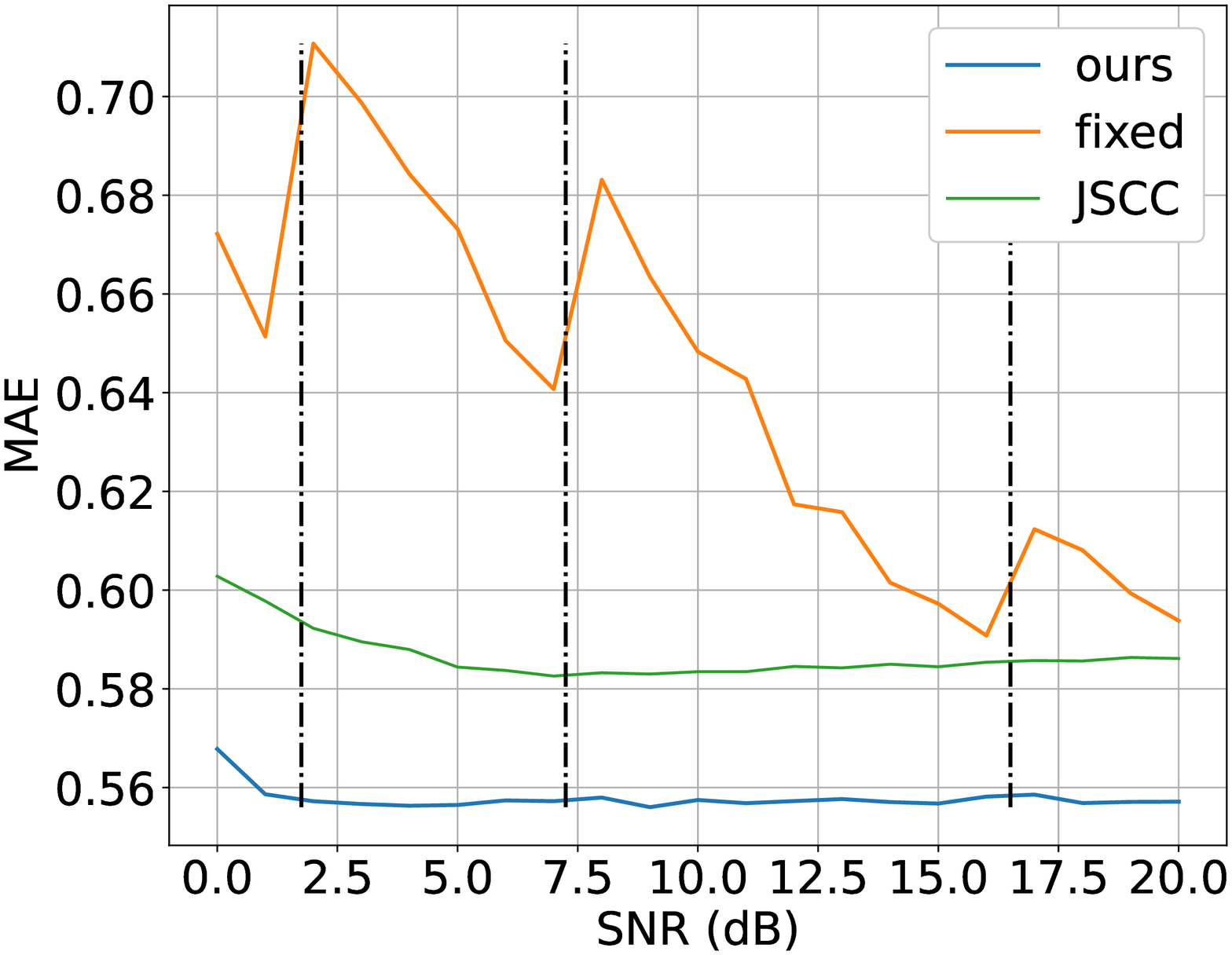}
            \end{minipage}
            \begin{minipage}[t]{0.45\textwidth}
            \centering
            \includegraphics[width=\textwidth]{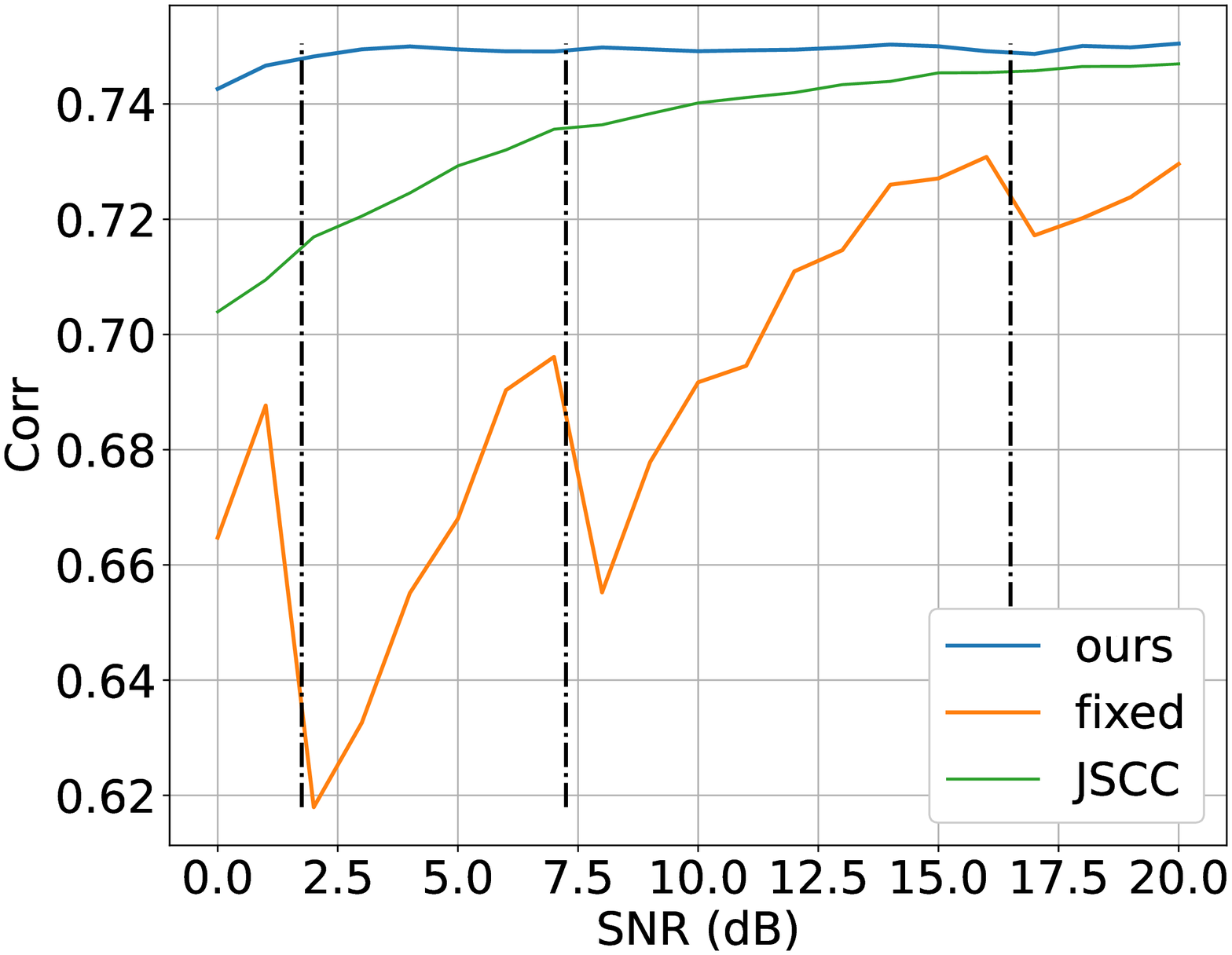}
            \end{minipage}
        }
    \caption{Performance comparison in the Rayleigh channel.}
    \label{fig:comparison_Rayleigh}
    \vspace{-0.5cm}
\end{figure}

To validate the superiority of our proposed rate-adaptive coding mechanism, we compare the performance of the distributed multi-modal fusion semantic communication framework with the aforementioned baseline methods. The results summarized in \autoref{fig:comparison_AWGN} and \autoref{fig:comparison_Rayleigh} illustrate that the proposed mechanism outperforms the baselines in terms of both semantic task accuracy and inference latency. 
Moreover, the performance of the proposed mechanism approaches the error-free transmission performance in both AWGN and Rayleigh channels, which indicates that it can indeed achieve robust semantic inference. The conventional method suffers from significant performance loss since the same transmission rate is applied for all modalities. To meet the delay requirement, a high transmission rate $R^{\text{fixed}}$ is adopted for all modalities. In this case, the distortion of semantically critical modality increases, and thus the robustness bound of the semantic task is exceeded. 

The performance deterioration denoted by the dash lines in \autoref{fig:comparison_AWGN} and \autoref{fig:comparison_Rayleigh} is due to modulation method selection. Intuitively, the transmission rate increases with the channel SNR and thus both higher modulation order and lower coding rate are chosen. As a result, the error protection ability decreases and the inference system suffers from performance degradation. From the figures, our transmission mechanism outperforms the NN-based JSCC approach. Clearly, the insertion of a channel layer in the original multi-modal fusion $\mathcal{F}$ makes the training process hard to converge. Meanwhile, the new model $\mathcal{F}^{\text{JSCC}}$ fails to reach the performance upper bound of the original model. Therefore, there exists a small performance gap between the JSCC approach and the proposed mechanism even in the high SNR scenario.

\subsubsection{Implementation Complexity}
\label{sec:complexity}

Finally, we study the implementation complexity to gain more insights into the proposed rate-adaptive coding mechanism in a realistic system. 
% In our simulation, the complexity of one layer LSTM is $\mathcal{O}(n\cdot d)$ \cite{LSTM} while the complexity of one layer Transformer is $\mathcal{O}(n^2\cdot d+n\cdot d^2)$ \cite{Transformer}, where $d$ is the feature dimension and $n$ is the sequence length. 
% Therefore, the complexity of the semantic source encoder and decoder can be computed, respectively as $\mathcal{O}(N_{\mathcal{S}})=\mathcal{O}(12(n^2\cdot d^{(1)}+n\cdot {d^{(1)}}^2) + 2(n\cdot d^{(2)} + n\cdot d^{(3)}))$ and $\mathcal{O}(N_\mathcal{C})=\mathcal{O}(9\cdot d+3\cdot d^2)$. Thus, the complexity of JSCC can be rewritten as
% \begin{equation}
%     \mathcal{O}(N_{\text{epoch}}\cdot N_{\text{sample}} \cdot (12(n^2\cdot d^{(1)}+n\cdot {d^{(1)}}^2) + 2n(d^{(2)} + d^{(3)}) + 9\cdot d+3\cdot d^2)).
% \end{equation}
In our simulation, the complexity of one layer Transformer is $\mathcal{O}\left(n^2\cdot d+n\cdot d^2\right)$ \cite{Transformer}, where $d$ is the feature dimension and $n$ is the sequence length.
Therefore, the complexity of the semantic source decoder can be computed as $\mathcal{O}\left(N_\mathcal{C}\right)=\mathcal{O}\left(9\cdot d+3\cdot d^2\right)$, and the complexity of JSCC can be rewritten as
\begin{equation}
    \mathcal{O}\left(N_{\text{epoch}}\cdot N_{\text{sample}} \cdot \left(9\cdot d+3\cdot d^2\right)\right).
\end{equation}
We take the model on the MOSI dataset as an example and approximately compute the number of basic operations as $28\times 1283 \times (9\times 128 + 3\times 128^2)=1.81\text{GOps}$, where GOps means billions of operations.

% MOSI样本数为1283，平均长度42，28epoch收敛
% MOSEI样本数16315，平均长度51，20epoch收敛

For the proposed rate-adaptive coding mechanism, we take MSA on the MOSI dataset as an example. The number of basic operations of the rate-adaptive coding mechanism can be approximately computed as $9\times 128 + 3\times 128^2-\log_2\left(10^{-6}\right)=56.62\text{MOps}$. We summarize the number of basic operations needed for deployment in \autoref{tab:operations}. It can be seen that the implementation complexity of the proposed rate-adaptive coding mechanism is much lower than JSCC. Specifically, for the MOSI dataset, the proposed mechanism simply requires $0.0028$\% number of operations of its counterpart to be deployed in a realistic system.

Based on the end-to-end NN, the JSCC approach could indeed mitigate the effect of channel noise. However, such an approach cannot directly use the existing pre-trained SOTA model. It requires re-train or fine-tune by adding a layer that simulates a wireless channel between the semantic encoder and decoder. Additionally, even if the JSCC approach has already designed and trained a new model with a layer of AWGN channel beforehand, it may not function optimally when the wireless environment changes, e.g. the distribution of channel fading changes. In such a case, the model must undergo re-training on the new wireless channel.

In contrast, the proposed rate-adaptive coding mechanism addresses this challenge through two approaches. First, the proposed mechanism only focuses on the inference process rather than the training process. To achieve the tradeoff between inference accuracy and delay, the proposed mechanism optimizes the transmission rates for all modalities based on both channel conditions and their semantic significance. As a result, the proposed mechanism can be applied to any new task with an existing pre-trained SOTA model without the need for re-training the model with a wireless channel layer. Second, the proposed mechanism employs conventional physical transmission methods, such as channel encoding and modulation, which are well-established in modern communication systems and applicable to most wireless environments. Furthermore, the optimization problem considers channel conditions based on CSI feedback, without being dependent on the type of wireless channel and its specific characteristics. This eliminates the need of re-training the model in a new wireless channel when the wireless environment changes.

% On the contrary, the proposed rate-adaptive coding mechanism utilizes an existing pre-trained  model for the semantic source encoder without extra re-training. The emphasis of the proposed mechanism lies in semantic-based unequal error protection in channel coding. As a result, the proposed mechanism could be deployed in well-established communication systems with considerably less effort compared with JSCC.

In summary, the JSCC approach jointly considers semantic information and wireless channel conditions by using the neural network, which requires training and lacks generality. On the contrary, the proposed mechanism considers semantic importance and channel conditions during the inference process by assigning unequal error protection to each modality based on RVP. This process is independent of the neural network's training process and is generally applicable to most neural network architectures and wireless channels. Therefore, the proposed mechanism could be deployed in well-established communication systems with considerably less effort compared with JSCC.

\begin{table}[htb]
    \caption{The number of basic operations needed for deployment.}
    \begin{center}
        \begin{tabular}{|m{0.2\textwidth}<{\centering}|m{0.2\textwidth}<{\centering}|m{0.2\textwidth}<{\centering}|}
            \hline
            Scheme  & JSCC  & Ours  \\
            \hline
            MOSI    & 1.81GOps & 56.62MOps  \\
            \hline
            MOSEI   & 16.41GOps & 56.62MOps  \\
            \hline
        \end{tabular}
        \label{tab:operations}
    \end{center}
    \vspace{-1cm}
\end{table}

\section{Conclusion}
\label{sec:conclusion}

This paper has presented a rate-adaptive coding mechanism in a distributed multi-modal semantic communication framework, which jointly considers the semantic source encoder and decoder with unequal protection channel coding. To balance the tradeoff between communication efficiency and semantic task performance, we modeled it as an optimization problem with guaranteed performance constraints. It aims to find the optimal transmission rate for each modality feature. We have analyzed the effect of channel coding rate on semantic task accuracy by utilizing the RVP on semantic encoder/decoder and constrained the optimization with a robustness bound of semantic output. The robustness bound is a function of modality features, channel condition, transmission rate, and semantic decoder, which makes the aforementioned optimization problem challenging to solve. Through approximation and relaxation, we have successfully solved the problem based on the optimization approach. Compared with the competitive semantic communication framework and conventional methods, the proposed mechanism achieves better semantic performance under the same inference delay. Additionally, the proposed mechanism can be generalized to any multi-modal fusion model and can be easily deployed without re-training the NN as compared to existing semantic communication frameworks.

As far as we know, we are among the first to investigate unequal protection for semantic communications with multi-modal data. This work has currently been a first step towards the goal and some open problems still need further study. 
First, this paper only focuses on the channel coding rate which is the most important error protection factor for channel codes. 
However, new designs of unequal error protection codes for multi-modal semantic communications that yield higher performance gain would be an intriguing direction to explore.
Secondly, we show that combining multiple independently faded replicas of semantic information would achieve more reliable inference. However, this performance gain is different from the one provided by diversity \mbox{\cite{diversity}}, and we name such a technique as "semantic diversity". It would be a promising future direction to theoretically analyze semantic diversity and combine semantic diversity and spatial diversity for higher performance gain.

% or

% \appendix[Convexity of $\mathcal{P}2$]


\appendices

\section{Proof of dual norm}
\label{apx:dual_norm}
Firstly, suppose $\|\cdot\|_p$ is an $\mathcal{L}_p$ norm on space $\mathcal{X}$, and $\boldsymbol{x}\in\mathcal{X}$. Let $\boldsymbol{y}=(y_1,\cdots, y_n)$ be an element of the dual space $\mathcal{Y}$. By definition, the dual space norm is given by
\begin{equation}
    \|\boldsymbol{y}\|_*=\sup_{\|\boldsymbol{x}\|_p\leq 1}\boldsymbol{y}^\top\boldsymbol{x}.
\end{equation}

By H$\ddot{\text{o}}$lder's inequality, we have
\begin{equation}
    \boldsymbol{y}^\top\boldsymbol{x} = \sum_{i=1}^n x_iy_i \leq \sum_{i=1}^n \left|x_iy_i\right| \leq \|\boldsymbol{x}\|_p\|\boldsymbol{y}\|_q,
\end{equation}
where $q = p/(p-1)$. Therefore, $\|\boldsymbol{y}\|_*\leq \|\boldsymbol{y}\|_q$.

To prove the converse inequality, we consider the vector defined by $\Tilde{x}_i=\frac{|y_i|^{q-2}}{\|\boldsymbol{y}\|_q^{q-1}}y_i$. Hence the $\mathcal{L}_p$ norm of $\Tilde{\boldsymbol{x}}$ is given by
\begin{equation}
    \|\Tilde{\boldsymbol{x}}\|_p = \frac{1}{\|\boldsymbol{y}\|_q^{q-1}}\left(\sum_{i=1}^n|y_i|^{p(q-1)}\right)^{1/p} = \frac{1}{\|\boldsymbol{y}\|_q^{q-1}}\left(\left(\sum_{i=1}^n|y_i|^{q}\right)^{1/q}\right)^{q/p} = \frac{\|\boldsymbol{y}\|_q^{q/p}}{\|\boldsymbol{y}\|_q^{q-1}} = 1.
\end{equation}
Since $\|\Tilde{\boldsymbol{x}}\|_p\leq 1$, the dual norm is the supremum over $\boldsymbol{y}^\top\Tilde{\boldsymbol{x}}$. Thus,
\begin{equation}
    \|\boldsymbol{y}\|_*\geq \boldsymbol{y}^\top\Tilde{\boldsymbol{x}}=\frac{1}{\|\boldsymbol{y}\|_q^{q-1}}\sum_{i=1}^n|y_i|^q = \|\boldsymbol{y}\|_q.
\end{equation}

Therefore, we have $\|\boldsymbol{y}\|_* = \|\boldsymbol{y}\|_q$ which proves that $\|\cdot\|_q$ is the dual norm of $\mathcal{L}_p$ norm.

\section{Convexity of $\mathcal{P}2$}

\label{apx:convex}
Apparently, both the objective function and the first constraint are convex. Thus $\mathcal{P}2$ is a convex optimization problem if the second constraint is also convex. We denote function $g$ as
\begin{equation}
    g\left(t,\left\{R^{(m)}\right\}_M\right) = \sum_{m=1}^M\frac{a^{(m)}}{1 + \text{exp}\left[k^{(m)}\left(b^{(m)}-R^{(m)}\right)\right]} - \Delta_0.
\end{equation}

Without loss of generality, we first compute the second order partial derivative with respect to $R^{(m)}$, as
\begin{equation}
    \label{equ:twice-derivatives}
    g_{R^{(m)}R^{(m)}} = \frac{\partial^2g}{\partial {R^{(m)}}^2} = a^{(m)}{k^{(m)}}^2\frac{\text{exp}\left[k^{(m)}(b^{(m)}-R^{(m)})\right]\left(\text{exp}\left[k^{(m)}\left(b^{(m)}-R^{(m)}\right)\right]-1\right)}{\left(1+\text{exp}\left[k^{(m)}\left(b^{(m)}-R^{(m)}\right)\right]\right)^3}.
\end{equation}

Since we have $b^{(m)} = C^{(m)} + \log L/L\geq R^{(m)}$ and $a^{(m)}\geq 0,\ \forall\ m=1,2,\cdots M$, the second order derivatives are non-negative, i.e. $g_{R^{(m)}R^{(m)}}\geq 0,\ \forall\ m=1,2,\cdots M$. In addition, it is clear that the second order partial derivatives regarding two different variables satisfy $g_{R^{(m)}R^{(n)}}=0,\  \forall\ m,n=1,2,\cdots M,\ m\neq n$. The Hessian of function $g$ given in \cref{equ:hessian} is positive semi-definite, which indicates that the second constraint is convex. Consequently, $\mathcal{P}2$ is a convex optimization problem.
\begin{equation}
    \label{equ:hessian}
    \mathbf{H} = \text{diag}(g_{R^{(1)}R^{(1)}},g_{R^{(2)}R^{(2)}},\cdots,g_{R^{(M)}R^{(M)}}).
\end{equation}

% % use section* for acknowledgment
% \section*{Acknowledgment}


\ifCLASSOPTIONcaptionsoff
  \newpage
\fi

\end{document}